**Synthesis pathways to thin films of stable layered nitrides**

Andriy Zakutayev[1*], Matthew Jankousky[2], Laszlo Wolf[2], Yi Feng[3], Christopher L. Rom[1], Sage R. Bauers[1], Olaf Borkiewicz[4], David A. LaVan[3], Rebecca W. Smaha[1], Vladan Stevanovic[2]

[1] National Renewable Energy Laboratory

[2] Colorado School of Mines

[3] National Institute of Standards and Technology

[4] Argonne National Laboratory

[*]andriy.zakutayev@nrel.gov

**Abstract**

Controlled synthesis of metastable materials away from equilibrium is of interest in materials chemistry. Thin film deposition methods with rapid condensation of vapor precursors can readily synthesize metastable phases, but often struggle to yield the thermodynamic ground state. Growing thermodynamically-stable structures using kinetically-limited synthesis methods in important for practical applications in electronics and energy conversion. Here, we reveal a synthesis pathway to thermodynamically-stable ordered layered ternary nitride materials, and discuss why disordered metastable intermediate phases tend to form. We show that starting from elemental vapor precursors leads to a 3D long-range disordered $MgMoN_2$ thin film metastable intermediate structure, with a layered short-range order that has a low-energy transformation barrier to the layered 2D-like stable structure. This synthesis approach is extended to $ScTaN_2$, $MgWN_2$ and $MgTa_2N_3$, and may lead to the synthesis of other layered nitride thin films with unique semiconducting and quantum properties.

## Introduction

Controlled synthesis of metastable solid-state materials away from thermodynamic equilibrium has been a significant challenge.[1] Synthesizing metastable polymeric, metallic, and semiconducting materials has important implications to electronic technologies and energy conversion. The reciprocal challenge is how to synthesize thermodynamically-stable structures using kinetically-limited thin film deposition methods, that are often used in manufacturing but tend to favor metastable material polymorphs. Among nitride thin films, examples of metastable thin film materials used for practical applications include MoN as an electrocatalyst for hydrogen evolution[2] and NbN in superconducting qubits,[3] but the corresponding thin film properties of their ground state structures remain largely unknown. An additional reason to examine thin film synthesis pathways to stable materials with layered 2D-like structures is that they tend to host unusual semiconductor or quantum properties promising for practical applications.[4] Among nitrides, examples include high mobility in 2D layered hexagonal boron nitride,[5] 2D electron gas at GaN/AlGaN interfaces[6], and diverse properties of nitrogen-containing intermetallic MAX phases[78] .

There are many known nitrides that are thermodynamically stable in layered, 2D-like crystal structures,[9] such as $ScTaN_2$,[10,11] $MgTa_2N_3$,[12,13] and $BaZrN_2$.[14,15] These materials belong to a layered family of multivalent ternary nitrides that combines two metal cations with a nitrogen anion in equal amounts and charge balanced stoichiometry.[16] These layered 2D-like materials feature alternating sheets of $MN_x$ polyhedra (where $M$ is a metal and $x$=4–10), such as octahedra, tetrahedra, square pyramids and trigonal prisms, and are distinct from 3D wurtzites[17] , 3D rocksalts,[18] or 3D perovskites[19] formed by elemental layering.[20] Several multivalent ternary nitrides with 1-1-2 stoichiometry ($FeWN_2$, $MnMoN_2$,[21] $CoMoN_2$[22]) and 1-2-3 stoichiometry ($MgTa_2N_3$,[12] $MgNb_2N_3$) have been synthesized in the layered structure by bulk methods, [23,24] and others ($ZnZrN_2$,[25] $ZnMoN_2$,[26] , $MgWN_2$[27]) have been calculated to be stable in 2D- layered structures.

Some layered 2D-like nitride materials are theoretically predicted to have unique properties, such as tunable topological and Dirac semimetal states in $MgTa_2N_3$ based alloys.[2829] Unfortunately, their electrical transport and optoelectronic properties remain experimentally unknown: these properties are difficult to measure for polycrystalline powders, and single crystals are even more challenging to grow in nitride chemistry. Thin films suitable for property measurements are commonly synthesized by rapid quenching from vapor or plasma to solid phase at a rate of up to $10^{10}$ K $s^{-1}$.[30] This synthesis pathway leads to metastable polymorphs with isotropic 3D bonding and very different properties compared to the thermodynamically-stable 2D- layered structures.[16] For instance, a layered 2D-like ground state of $MgWN_2$ is a semiconductor, whereas a 3D metastable polymorph of $MgWN_2$ is a metal.[27] It remains unclear both i) why thin film synthesis favors metastable 3D structures, and ii) how to reliably achieve nitrides with layered 2D-like structures in film form.

Here we describe a synthesis pathway to nitride thin films with stable layered 2D-like crystal structures, though a 3D long-range disordered metastable intermediate structure. While growth on heated substrates results in decomposition above 600 °C, the $MgMoN_2$ with a stable layered 2D-like cation-ordered structure forms after 700 °C – 1100 °C annealing of a 3D metastable intermediate structure deposited at ambient temperature. The emergence of this cation-ordered layered 2D-like product may seem surprising because the 3D isotropic RS structure is apparently cation-disordered according to X-ray diffraction (XRD), despite the difference in Mg and Mo atom properties. Experimental and theoretical pair distribution function (PDF) analysis reveals a layered short-range order in the RS structure, that is reminiscent of the polyhedral layering within the RL structure. We theoretically identify a simple displacive transformation pathway from a 3D RS intermediate to a layered 2D-like RL product, and experimentally quantify a 100 meV/atom transformation barrier consistent with the calculations. These $MgMoN_2$ results give insight into synthesis of layered 2D-like products, and are extended here to $ScTaN_2$, $MgWN_2$ and $MgTa_2N_3$, which may aid the synthesis of many other layered materials in thin film form.

## Results and Discussion

### *Experimental discovery of rocksalt-to-rockseline transition*

Crystal structures of the layered 2D-like multivalent ternary nitride materials are most easily described as a mixture of the 3D structure archetypes that embody each layer (Fig. 1). For example, $ZnZrN_2$ and $ZnHfN_2$ are predicted to form in thermodynamically-stable layered 2D-like "wurtsalt" (wurtzite–rocksalt, or WS; see Fig. 1a) structure,[31] but thin film synthesis of $ZnZrN_2$ up to 300 °C results in metastable 3D RS or h-BN crystal structures[25]. Similarly, the $Zn_3MoN_4$–$ZnMoN_2$ family crystallizes in a 3D wurtzite structure across the entire chemical composition range, instead of crystallizing in the thermodynamically-stable layered 2D-like "wurtzeline" (wurtzite–nickeline, or WL; see Fig. 1b) structure at $ZnMoN_2$ composition.[26] Recently, $MgWN_2$ thin film synthesis on a heated substrate also resulted in two metastable 3D RS and h-BN structures, whereas the stable layered 2D-like "rockseline" (rocksalt–nickeline, or RL; see Fig. 1c) structure with 1-1-2 stoichiometry was synthesized by bulk ceramic methods,[27] and rockseline structures with 1-2-3 stoichiometries have been also reported for $MgTa_2N_3$ [12](Fig. 1d). Growing these Zn- and Mg-based ternary thin films on substrates held at higher temperature to achieve the ground state layered structure was not feasible, likely due to re-evaporation of $ZnN_x$- and $MgN_x$-based components caused by their high vapor pressure.

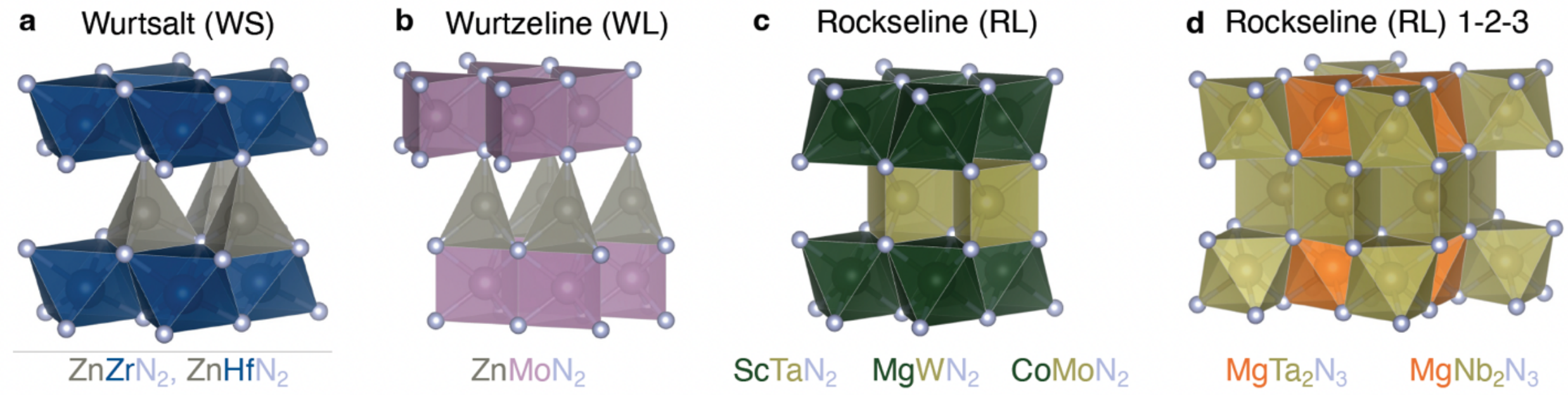

**Figure 1: Examples of layered, 2D-like crystal structures among multivalent ternary nitrides**: (a) "wurtsalt" wurtzite–rocksalt, or WS, structure consisting of alternating sheets of tetrahedra and octahedra. (b) "wurtzeline" wurtzite–nickeline, or WL, structure consisting of alternating sheets of tetrahedra and trigonal prisms. (c) "rockseline" rocksalt–nickeline, or RL, structure with 1-1-2 stoichiometry consisting of alternating sheets of edge-sharing octahedra and edge-sharing trigonal prisms; and (d) RL structure with 1-2-3 stoichiometry.

To experimentally realize $MgMoN_2$ with layered 2D-like crystal structure, thin films of Mg–Mo–N were synthesized by radio-frequency (RF) co-sputtering from metallic Mg and Mo precursors in Ar and $N_2$ atmosphere, and then subjected to rapid thermal annealing under atmospheric $N_2$ pressure. The results of composition measurements indicate that the synthesized $MgMoN_2$ materials in the Mg/(Mg+Mo) = 0.50 – 0.75 range (as measured by XRF) are free of oxygen impurities (Fig. 2a, Fig. S1a). According to X-ray diffraction (XRD) results, the as-deposited Mg–Mo–N films form in a nanocrystalline RS-derived crystal structure with mixed Mg and Mo occupancy of the cation lattice across a broad range of studied compositions (Mg/(Mg+Mo) = 0.20 – 0.95) and temperatures (Fig. 2b). LeBail fits of the RS lattice constants as a function of cation composition in the as-deposited samples measured by synchrotron grazing incidence XRD show the expected linear trend (Fig. 2c).

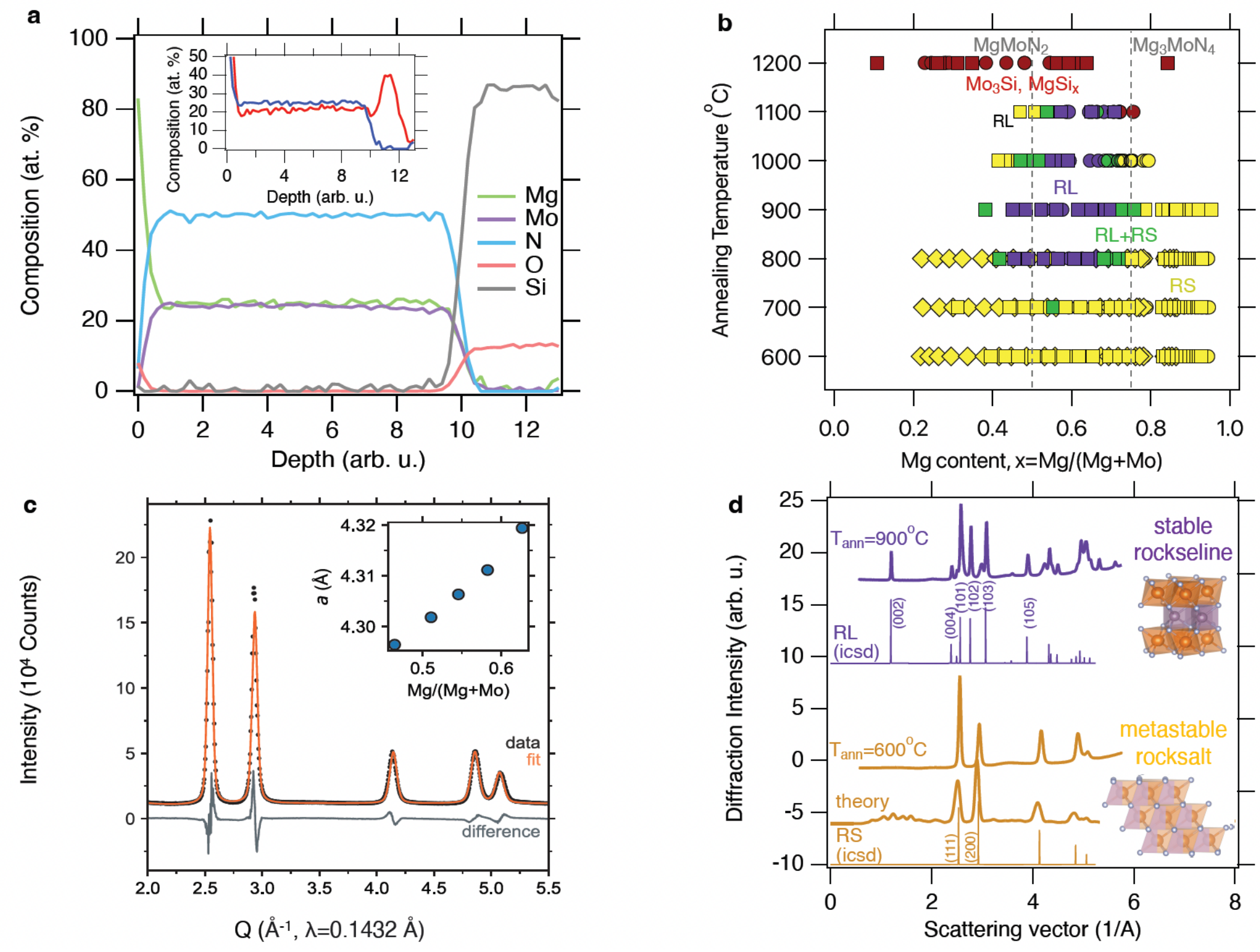

**Figure 2: Chemical composition and crystal structure of $MgMoN_2$:** (a) Composition measurements as a function of depth showing negligible amount of oxygen, and some Mg accumulation at the substrate interface after annealing (inset). (b) Growth map showing formation of the layered 2D-like RL phase at $MgMoN_2$ and Mg-rich compositions above 700 °C. (c) LeBail fit performed on synchrotron grazing incidence XRD in the RS structure for as-deposited films, and the relationship between lattice parameter *a* and cation ratio (inset). (d) Synchrotron GIWAXS data and crystal structures of $MgMoN_2$ in metastable 3D RS and stable layered 2D-like RL phases, along with two theoretical reference patterns from ICSD, and one computationally simulated average disordered RS pattern

After annealing for 3 min at $T_{ann}$ = 800 – 1100 °C, the crystal structure transforms from RS to RL (see Fig. 2c and Fig. 2d), with phase-pure RL material obtained in the Mg/(Mg+Mo) = 0.45 – 0.70 range. This layered phase cannot be achieved by depositing on the substrate which is heated to 800 °C under low sputtering pressure because even at 600 °C, there is very little Mg incorporation in the growing film due to high $MgN_x$ vapor pressure (Fig. S2a), similar to $MgWN_2$[27] and $ZnN_x$ in $ZnZrN_2$.[25] Outside of the Mg/(Mg+Mo) = 0.45 – 0.70 composition window, the RS structure persists after annealing. However, when annealing above 1100 °C (Fig. 2b) and for longer than 3 min (Fig. S1b) a significant reaction of Mo with the Si substrate is observed, as indicated by $Mo_3Si$ diffraction peaks.

The results of synchrotron grazing incidence wide-angle X-ray scattering (GIWAXS) measurements for $MgMoN_2$ films annealed at 600 °C (RS phase) and 900 °C (RL phase) are shown in Fig. 2d alongside the two crystal structure models. All RS and RL peaks expected from reference patterns are clearly resolved, with some deviation in the peak intensity due to preferential orientation effects or Mg/Mo cation disorder across $O_h/T_p$ sites. GIWAXS patterns for the $Mg_3MoN_4$ composition with the RS structure are shown in Fig. S2b. The measured RL $MgMoN_2$ data show a strong low-angle (002) and subsequent (00*L*) peaks indicative of some degree of Mg and Mo ordering on the $O_h$ and $T_p$ sites in the RL structure, whereas the measured RS $MgMoN_2$ data do not show any peaks related to this long-range cation ordering. It is surprising that the apparently cation-disordered RS structure transitions into a clearly cation-ordered RL structure during the annealing process. and that the metastable 3D RS structure forms in the first place instead of the stable layered 2D-like RL structure.

***Theoretical explanation of metastable rocksalt synthesis***

From a theoretical point of view, the experimental realizability of various crystalline phases is related to the size of the corresponding local minima on the potential energy surface. The size of a local minimum – the total volume occupied by its attraction basin in the configuration space of atoms – measures the probability to of a structure to fall into that minimum. Hence, a larger basin of attraction implies higher chances for its experimental realization. As was shown previously, this principle explains all experimentally observed metastable polymorphs in several elemental or binary chemistries, such as elemental silicon,[32] binary oxides,[33] and group-IV carbides,[34] and the emergence of the disordered RS phase in the thin film synthesis of $ZnZrN_2$ in lieu of the ordered "wurtsalt" ground state.[25]

To explain the $MgMoN_2$ experimental results described above, we investigated relevant features of the potential energy surface (PES) of $MgMoN_2$ using the first-principles structure sampling [33]. This involved generating 5,000 random structures, relaxing them to the closest local minimum, and classifying them into groups with the same underlying structure-type (see Methods). As shown in Fig. 3a, the tallest peak in the space group-resolved thermodynamic density of states (TDOS) with 314 members has the RS-type parent structure with space group (s.g.) 225 (*Fm*-3*m*), with the relatively narrow energy range of 0.5 – 1.5 eV/f.u. due to the variations in the Mg/Mo site occupancies, like in entropy-stabilized ceramics. There are two other relatively populous 6-fold coordinated hexagonal structures (s.g.160 and s.g.166) with 10x – 50x smaller representation than RS, made of edge- and face-sharing octahedra (s.g.166) or a combination of octahedra and trigonal prisms (s.g.160), which can be viewed as a disordered version of the ground state s.g.194 (shown as a dashed line in Fig. 3a). The low symmetry structure-types (s.g.1 and s.g.2) have high energy and large spread, so they are important for description of the amorphous or glassy state.[35]

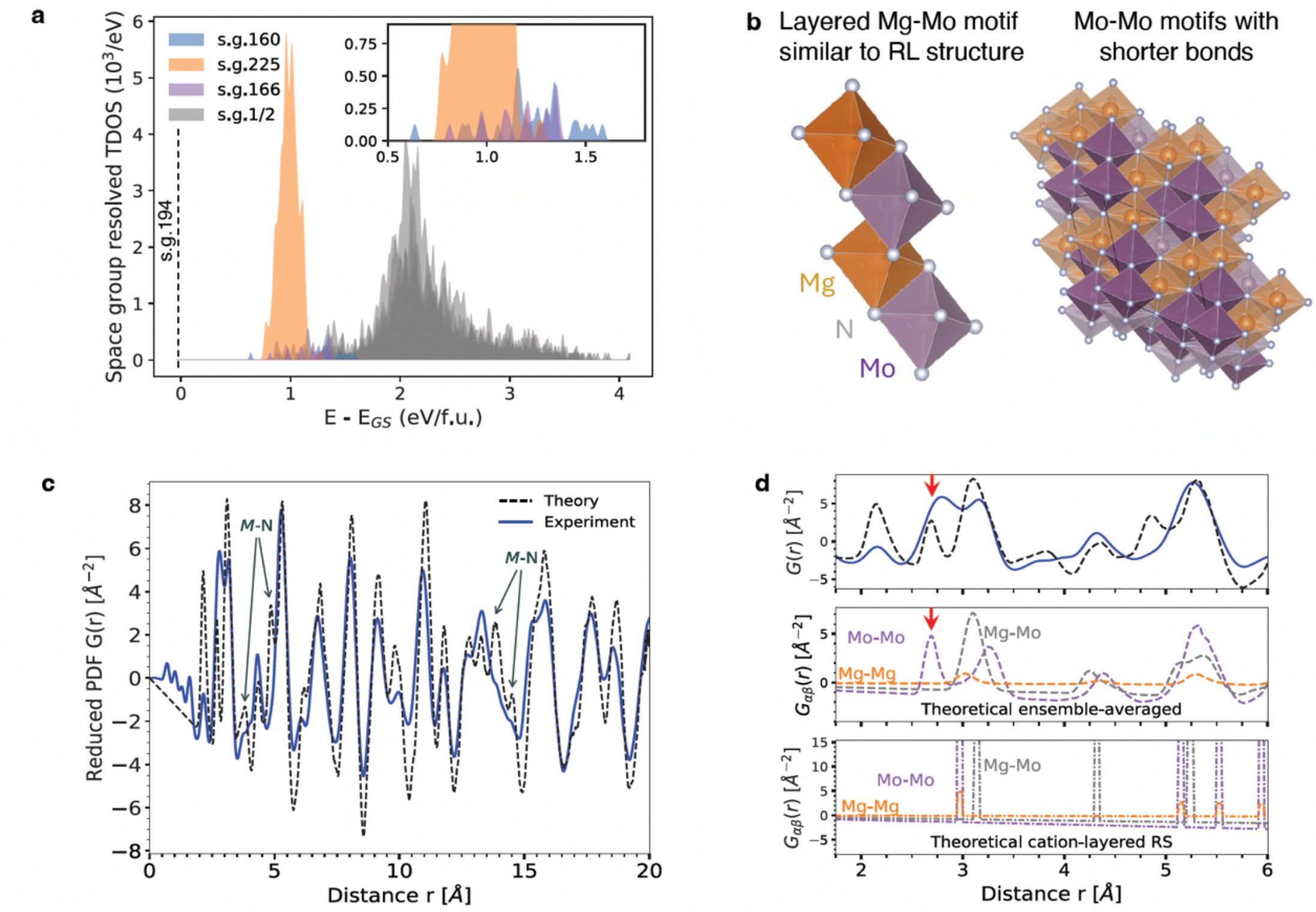

**Figure 3: Local short-range atomic order in metastable RS $MgMoN_2$** (a) Space group-resolved thermodynamic density of states (TDOS) of $MgMoN_2$, showing the prevalence of the RS structure s.g.225. (b) The cation-layered -Mg-Mo-Mg-Mo RS motif along (111) axis (left), taken from a RS structure predicted by structure sampling (right), with shortened Mo–Mo distances (2.62 Å, 2.75 Å, 2.72 Å) in the highlighted $MoN_6$ octahedra. (c) Measured reduced PDF G(*r*) of as-deposited $MgMoN_2$ films (blue line) and theoretical model from structure sampling (dashed black line), showing good agreement. (d) Magnified comparison of experimental and theoretical PDF in the low-*r* region (top), with a red arrow indicating the 2.7 Å peak that does not exist in the cation-disordered RS model (bottom), but results from Mo–Mo interactions (middle).

Fig. S3a displays the temperature dependence of the configurational free energies of all structure-types from the structure sampling.[23] Our calculations include the RL phase, which is represented by the dashed line and assumed to have no configurational entropy (occurrence of 1). The RS phase of $MgMoN_2$ has the highest entropy because of its large and narrow TDOS, and it becomes the lowest free energy structure for temperatures above 2000 K (Fig. S3b). The as-deposited films at ambient conditions exhibit a cation-disordered RS structure predicted above 2000 K because their kinetically frozen cation disorder corresponds to this effective temperature. It has been shown that equilibrium calculations for 1000 K – 2500 K temperatures reproduce the cation disorder that exists in materials grown in thin-film form at non-equilibrium conditions, such as $ZnZrN_2$,[25] $Co_2ZnO_4$, and $Co_2NiO_4$.[36]

The RL phase known from bulk synthesis of $MgMoN_2$ (s.g.194)[23,24] did not occur a single time in the random structure sampling, indicating that the stable layered RL structure of $MgMoN_2$ has a very narrow local minimum on the PES. This is a markedly different result than for $ZnZrN_2$ and for all other systems we investigated, for which the ground state structures are consistently found by the structure sampling. However, atomic motifs having a -Mg-Mo-Mg-Mo- cation ordering, referred to as the "cation-layered RS motif," were identified within all RS-type structures from the structure sampling (Fig. 3b). In these motifs, Mo(1) is in the [110] crystallographic direction from Mg(1), Mg(2) is in the [101] direction from Mo(1), and Mo(2) is in the [110] crystallographic direction from Mg(2). Periodic boundary conditions applied to this motif result in the (111) superlattice with alternating Mg and Mo layers.

This four-atom -Mg-Mo-Mg-Mo- motif would be statistically probable to find even in a fully random alloy because the probability of finding a single sequence of four atoms occupied in a certain way is $(1/2)^4$. Moreover, given the relatively large number of ways a given motif can be constructed, a large fraction of cations can be expected to be a part of at least one such motif. Therefore, we hypothesize that the long-range ordered RL phase is experimentally realized (Fig. 2) by a facile structural transformation from the short-range ordered RS phase (Fig. 3b). For this to happen at the relatively low annealing temperature of 700 °C, the experimental RS structure should contain domains where the local arrangement of Mg and Mo atoms provides a seed for alternating (111) layers, similar to the alternating (001) planes of the RL product.

***Local structure from theoretical and experimental PDF***

The short-range cation order in the long-range disordered RS intermediate structure was measured using synchrotron grazing incidence PDF on a series of pre-annealed thicker (600 – 800 nm) $MgMoN_2$ samples, with the quartz substrate contribution measured and subtracted. The results of composition-dependent XRD measurements and LeBail fits (Fig. S4b) are consistent with the lattice constant linearly increasing as a function of Mg composition (Fig. 2c). The measured PDF was fit using a randomly disordered RS model across a wide $r$-range (2 – 30 Å, see Fig. S4a). While the high-$r$ range (> 10 Å) matched this model well, a surprising difference is the short-distance local ordering range (2 – 7 Å) where the PDF fit residuals are high, especially in the 2.5 – 3.0 Å region. This implies that the length scale of the short-range correlations is on the order of 5 Å, that would be hidden from traditional long-range XRD measurements. The lowest-$r$ (<2 Å) oscillations of the experimental data are due to truncated Fourier transform effects and substrate correction.

To understand the difference(s) between the experimentally measured PDF and the disordered rocksalt PDF model in the 2.0 – 3.0 Å region, a theoretical PDF was obtained through ensemble averaging over all RS-type structures (Fig. 3a). The positions of peaks in the experimental data are well matched by the theoretical model across the entire $r$-region (Fig. 3c), including the 2 – 3 Å range (Fig. 3d, top panel). Fig. 3d (middle and bottom panels) shows the contributions of specific atom type pairs to the PDF (denoted as the partial PDF). In the cation-layered RS structure, the nearest-neighbor Mo–Mo and Mg–Mg partial peaks are superimposed at the same distance $r$ = 3.1 Å (Fig. 3, bottom). In the ensemble-averaged theoretical PDF of the disordered RS model (Fig. 3d, middle), the Mg–Mo and the Mg–Mg partial peaks are also present at similar distances ($r$ = 2.95 Å). However the Mo–Mo partial is split into two peaks: a peak blending with the Mg–Mo peak at $r$ = 3.2 Å, and most importantly another peak around $r$ = 2.7 Å , also known in other layered ternary transition metal nitrides.[10] We identify that this splitting arises from the relaxations of the Mo atoms towards shorter distances that occur in disordered RS-type structures (Fig. 3d, middle).

The modeled Mo–Mo peak (Fig. 3d, middle) explains the measured PDF peak at approximately $r$ = 2.0 – 3.0 Å (Fig. 3d, top) that is not captured by a cation-layered (Fig. 3d, bottom) cation-disordered (Fig. S4a) RS models. It is unlikely that this measured peak is due to Mo–Mo distances in the metallic bcc-Mo structure, because its intensity scales with Mg content (Fig. S4b) and because the single-phase RS XRD LeBail fit residuals are low (Fig. S3b). The overall agreement between experimental and theoretical PDFs (Fig. 3c), and especially the agreement for the distinct Mo-Mo peak in 2.5-3.0 Å range, are the evidence for the similarity between the as-grown $MgMoN_2$ thin films and the ensemble average of RS-type structure samples (Fig. 3a), but not the RS model with random cation disorder (Fig. S4a). Due to this direct correspondence, the cation-layered -Mg-Mo-Mg- motif found in the RS-type samples structures (Fig. 3b), is theorized to also be found in the as-grown $MgMoN_2$ thin films. For this cation-layered RS motif (Fig. 3b), to transform to the site-layered RL structure, the kinetic barrier between these two phases should be low, as addressed next.

***Measurements and calculations of the transformation pathway***

To characterize the RS-to-RL crystallographic transformation process, we performed temperature-dependent XRD measurements on select samples at the $MgMoN_2$ composition. The XRD color intensity map as a function of measurement temperature (Fig. 4a) shows a sharp transition around 700 – 720 °C, suggesting a facile transformation pathway between RS and RL structures. The unusual negative thermal expansion of the RS lattice in the 300 – 600 °C range (Fig. S5) should also be noted. Time-dependent XRD measurements at two different temperatures in this range show a gradual increase of the RS (111) peak intensity and decrease of the RL (012) peak intensity (Fig. S6). Fitting the XRD data measured at 700 °C and 712 °C to a sigmoidal Avrami equation[37] in both cases leads to an exponent value of approximately $n$ = 1.7 (Fig. 4b). The measured $n$ = 1.7 exponent could correspond to a combination of 2D interface-controlled growth ($n_g$ = 2) that could be rationalized by the 2D thin film geometry, and a 3D diffusion-controlled growth ($n_g$ = 1.5) that could be rationalized by the observation of the long-range Mg–Mo atom diffusion (Fig. 2a)

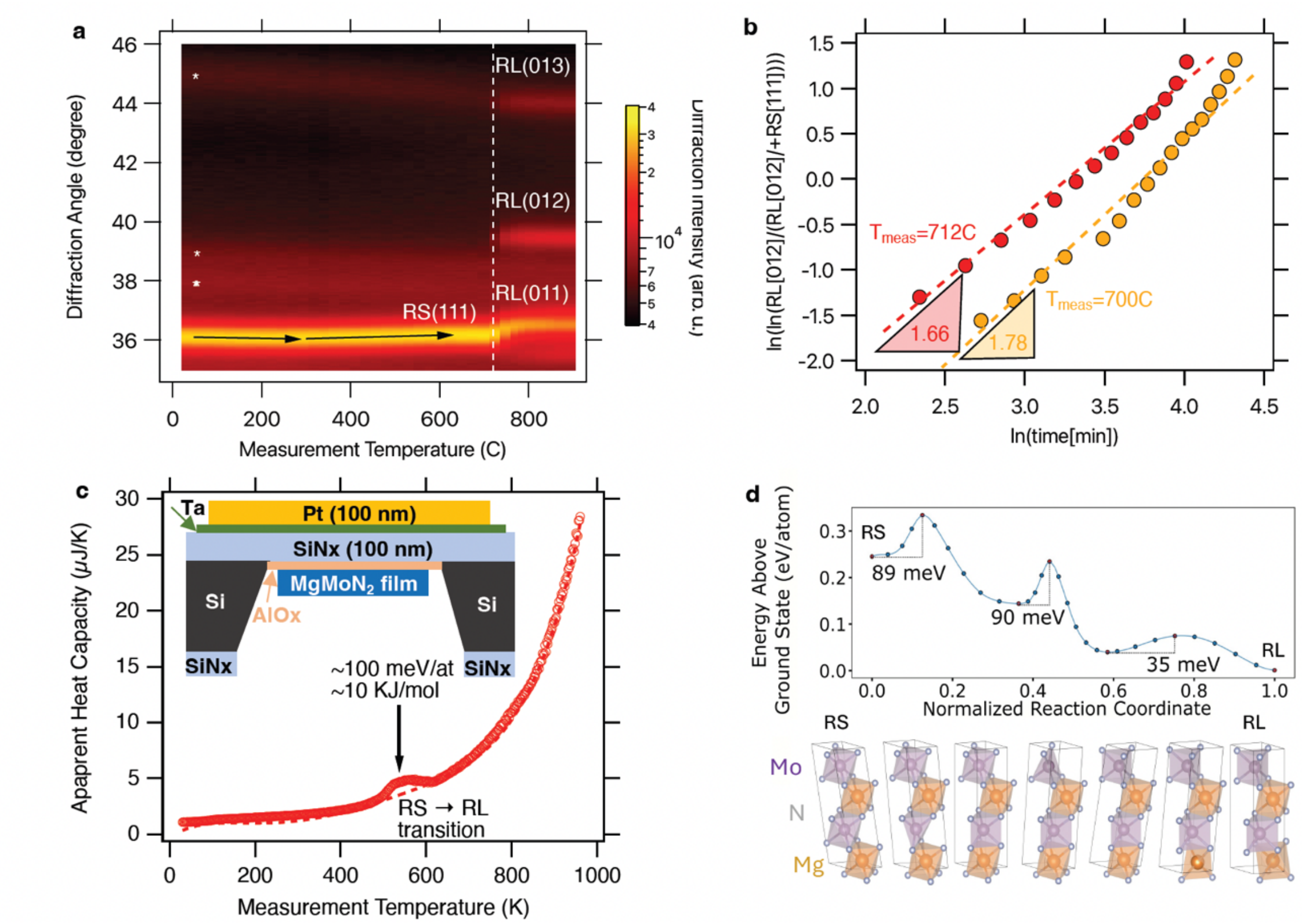

**Figure 4: In-situ measurements of the $MgMoN_2$ RS-RL phase transformation.** (a) Temperature-dependent XRD showing negative thermal expansion of RS phase in the 300 °C – 600 °C range and a sharp transition to RL phase around 700 °C. Asterisks indicate broad peaks from a polymeric sample cover dome. (b) Analysis of these time-dependent measurements at 700 °C and 712 °C, resulting in a slope of n=1.7,

which reflects the combination of instantaneous 2D nucleation (n=1.5) and 3D diffusion (n=2.0). (c) Nanocalorimetry results using measurement chip (inset), showing a small endothermic signal associated with overcoming the RS-RL energy barrier. (d) The energy profile (top) and the corresponding structure (bottom) for the transformation of the layered RS structure to the RL structure (continuous line), as calculated using SS-NEB (discrete symbols), with red symbols indicating the critical steps of the pathway (see text).

To measure the energy scale of the phase transformation process, we performed nanocalorimetry measurements in UHP $N_2$[38] for 100 nm thin $MgMoN_2$ films deposited on microfabricated Si chips. The apparent heat capacity as a function of measurement temperature (Fig. 4c) shows an endothermic signal that is 25 °C in width and approximately 10 KJ/mol (100 meV/at) in the integrated energy value. It is not entirely clear why this endotherm appears at 550 – 600 °C instead of at 700 °C, the temperature at which the phase transformation is observed by both ex-situ XRD (Fig. 2) and in-situ XRD (Fig. 4a), with possible temperature measurement error accounting for 25 – 75 °C. Possible physical reasons include (a) the need for growth of the locally transformed nuclei to be observed by extended crystal structure measurement methods, and (b) strong dependence of the phase transformation process on the heating rate (0.1 – 10 K/s for XRD vs. $10^3$ – $10^4$ K/s for nanocalorimetry). Nevertheless, the precise measurement of the small 100 meV/at calorimetry signal for a 100 nm nitride thin film is indicative of a relatively low kinetic barrier of the phase transformation process.

To calculate the energy barrier, we modeled the transformation pathway from the cation-ordered RS structure (α-$NaFeO_2$-type) to the cation-ordered 2D-like RL structure.[39] The calculated barrier of approximately 90 meV/atom (Fig. 4d) is consistent with the experimental observation that the transition occurs at approximately 600 – 700 °C, and the apparent change in heat capacity during that transition, suggesting that small domains of the cation-layered RS motif are present in the thin film. The first step of this pathway involves the transformation of layers of $MoN_6$ octahedra to $MoN_6$ trigonal prisms and is characterized by the relative shift of a layer of N atoms: one Mo–N bond per Mo atom is broken and then reformed in this process (89-90 meV/atom). In the second step, the layers of $MgN_6$ octahedra must invert: one Mg atom per two formula units must moves in every other layer, where two Mg–N bonds per Mg atom are broken (35 meV/atom). The barrier is not expected to change significantly with the introduction of cation disorder to the material, as shown for the barriers in supercells with Mg and Mo sites swapped (Fig. S7), though the 90 meV/atom rate limiting step switched from first to the second one.

***Discussion of rockseline $MgMoN_2$ crystallization mechanisms***

In this work, we experimentally discovered (Figs. 2, 4) and theoretically explained (Figs. 3, 4) a synthesis pathway to nitride thin films with layered 2D-like crystal structures (Fig. 1) from 3D atomically mixed intermediates deposited from vapor precursors. The key insights are that:

(i) The 3D structures are high-probability metastable phases favored by kinetically limited growth methods (Fig. 2, 3), whereas stable layered 2D-like structures have low probability of formation.

(ii) A layered 2D-like crystal structure can form by facile atomic transformation (Fig. 4) from a 3D intermediate that has no long-range cation order but features layered cation-ordered local motifs (Fig. 3).

(iii) These motifs (Fig. 3) serve as nucleation centers for this crystallographic transformation into a long-range ordered 2D-like structure experimentally achieved by annealing (Fig. 2, 4).

This synthetic pathway is supported by direct comparison of measured and calculated PDF (Fig. 3) and the energy barrier (Fig. 4) between the 3D intermediate and the 2D-like product.

A schematic representation of the synthetic pathway on an energy scale in the Mg–Mo–N compositional coordinates is shown in Fig. 5a. The first step of the process is the formation of a high-energy, long-range disordered but short-range ordered RS-$MgMoN_2$ 3D crystal structure from vapor phase Mg, Mo, and N precursors (Fig. 5a). The second step is transformation of this 3D metastable intermediate structure into a stable layered 2D-like RL-$MgMoN_2$ structure with both short-range and long-range cation order (Fig. 5a). Fig. 5b illustrates the potential energy surface at a fixed $MgMoN_2$ composition in energy vs. configurational space coordinates. The potential well corresponding to the ground-state layered RL-$MgMoN_2$ structure is deep and narrow. However, it is separated by a small 100 meV potential energy barrier from a much wider—and hence much more probable—3D RS-$MgMoN_2$ well (Fig. 5b) that is slightly higher in energy. This small potential barrier corresponds to the facile RS-RL local transformation pathway identified in Fig. 4.

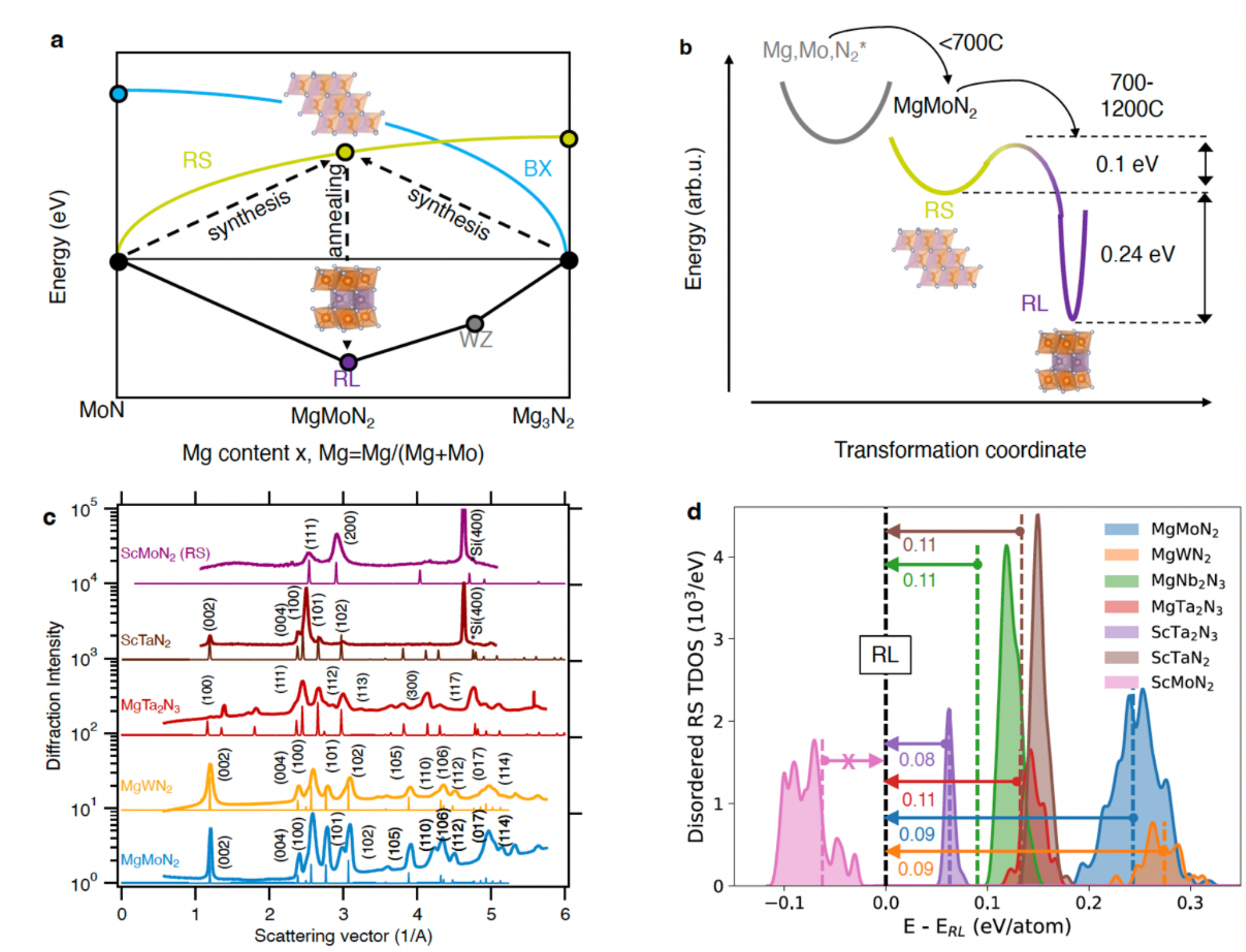

**Figure 5: 3D-to-2D synthesis pathway for $MgMoN_2$ and extension to other materials chemistries**: (a) Convex hull diagram, showing co-deposition of a high-energy 3D RS $MgMoN_2$ intermediate (yellow) and its crystallographic transformation to the layered 2D-like RL product (purple), with BX (blue) indicating the competing bixbyite structure of $Mg_3N_2$. (b) Potential energy landscape at a fixed $MgMoN_2$ composition, showing a low-energy 0.1 eV barrier between the broad metastable 3D RS energy valley (yellow) and a stable deep, narrow 2D-like RL energy well (purple). (c) Experimentally-measured synchrotron GIWAXS patterns of four layered 2D-like RL thin film nitride materials $MgMoN_2$, $MgWN_2$, $MgTa_2N_3$, and $ScTaN_2$ and one 3D RS $ScMoN_2$, including their minimal RS-to-RL annealing temperatures. (d) Calculated thermodynamic density of states and transformation energy barriers from the cation-layered RS to structurally layered RL, indicating that all of them except $ScMoN_2$ should transition from RS to RL structure.

The synthetic pathway reported here is not limited to thin film synthesis methods; it should also apply to bulk synthesis approaches where intimate atomic intermixing is involved. For example, in the case of $MgMoN_2$ synthesized in bulk by metathesis reaction, an intermediate 300 − 450 °C heating step to promote Mg−*M*−N bond formation, followed by 800 − 900 °C annealing of these intermediates, successfully resulted in the layered 2D-like RL-$MgMoN_2$ structure. In contrast, the direct heating to the 800 − 900 °C reaction temperature results in formation of a 3D RS structure, likely accompanied by volatilization of Mg-N.[40] Bulk syntheses of $MgMoN_2$[23] and $MgWN_2$[27] by conventional one-step route requires much higher temperature (1100 °C) and results in residual W or Mo and other impurities. In the case of $ScTaN_2$ bulk powders, heating to 1546 °C for 200 hours was required to obtain a phase-pure product,[10] illustrating the advantage of the facile transformation pathway to synthesis of layered ternary nitrides reported here.

***Extension to other layered multivalent ternary materials***

To demonstrate the universality of the synthesis pathway discovered in this work for RL $MgMoN_2$, we attempted synthesis of five other materials with alternating layers of Mg or Sc in $O_h$ coordination and Mo, W, or Ta in $T_p$ coordination. These ternary nitride materials include $MgWN_2$,[27] $MgTa_2N_3$,[13] $ScTaN_2$,[10] which have been reported to have a layered 2D-like crystal structure based on bulk synthesis, and $MgNb_2N_3$, $ScMoN_2$ that have been reported in the ICSD. Thin films of these materials deposited at ambient temperature crystallizes in a 3D disordered RS structure. The $MgWN_2$, $MgTa_2N_3$, $MgMoN_2$, $ScTaN_2$ annealed at 800 – 1200 °C (Fig. 5c) transform into layered 2D-like structures with cation ordering indicated by characteristic low-angle diffraction peaks. On the other hand, $ScMoN_2$ could not be synthesized in the RL structure (Fig. 5c), and the $MgNb_2N_3$ synthesis results were inconclusive (Fig. S8).

Fig. 5d shows the structure sampling (2,000 structures) and kinetic barrier modeling results for seven ternary nitrides including $MgWN_2$, $MgTa_2N_3$, $ScTaN_2$, $MgNb_2N_3$, $ScTa_2N_3$, $ScMoN_2$, $MgMoN_2$, in comparison with total energy calculations for RL structure (calculated for all prototypical structures from Fig. 1). The frequency of RS occurrence far exceeds any other structure type (see Fig. S9), and a high percentage of atoms participates in the RS layered motifs (Table S1). The calculated barriers for the structural transformation from the ordered RS into the ordered RL phase are all in the 0.08 – 0.11 eV/atom range (Fig. S10). These barriers correlate well with the annealing temperatures needed for the realization of a RL phase (Fig. S11). For all the investigated nitrides the RL structure has the lowest energy, with exception of $ScMoN_2$, indicating that the transformation from the RS structure is unlikely. This theoretical result can be explained by the $O_h$ rather than $T_p$ coordination preference of the Mo(III) ion with $d^3$ electronic configuration[41], and is consistent with the experimental observation (Fig. 5c) providing a false negative validation for our theoretical method.

The successful synthesis of $MgWN_2$, $MgTa_2N_3$, $MgMoN_2$, and $ScTaN_2$ layered nitride materials in thin film form (Fig.5c) and the theoretical explanation of their realization (Fig.5d), support generality of the transformation pathway described in this paper. These results are also interesting because thermoelectric[11] and quantum[29, 28] properties were predicted for these RL materials by theoretical calculations but have not been measured yet due to prior difficulty in thin film synthesis. The general synthesis pathway demonstrated in this work for layered RL structure may also apply to nitrides with $KCoO_2$-type structures such as $SrTiN_2$ and $BaZrN_2$ (Fig. S12), with two distinct coordination environments that recently attracted renewed attention due to their electronic properties.[42,43] This pathway may also apply other site-layered ternary nitrides, but likely not to cation-layered single-site wurtzite[17] or rocksalt[18] phases. It would be also interesting to perform similar calculations and experiments for the layered nitride and carbide MAX phases[8], where only 4 – 6 out of >160 materials have been grown in thin film form.[7]

**Conclusions**

We synthesized $MgMoN_2$ with a stable layered 2D-like crystal structure from a 3D disordered metastable intermediate, and quantified the 3D-to-2D transformation pathway by in-situ measurements and theoretical calculations. As-deposited $MgMoN_2$ thin films crystallize in an apparently cation-disordered rocksalt (RS) structure with 3D octahedral coordination, and rapidly transforms into a cation-ordered layered 2D-like RL structure above 700 °C. Experimental PDF measurements show peak splitting around 2.7 A, supporting the computational model for short-range ordered 3D RS structure. The calculated transformation pathway has a 100 meV/atom energy barrier measured using nanocalorimetry, further validating the 3D-to-2D transformation model. This $MgMoN_2$ discovery means that other metastable 3D nitride materials may have short-range order and transformation pathway into to stable layered 2D-like crystal structures, such as $MgWN_2$, $MgTa_2N_3$, and $ScTaN_2$ shown here. These results also imply that the long-range order of the final layered product should be possible to control through altering short-range order of the intermediate during the synthesis, which is important for fine-tuning materials quantum or semiconducting properties.

## Methods

### ***Experimental***

We synthesized thin films (typically 150 – 300 nm thick) of Mg–Mo–N by radio-frequency (RF) co-sputtering from metallic Mg and Mo precursors in an Ar and $N_2$ atmosphere onto Si substrates without (80 °C) and with (<600 °C) intentional heating, and subjected them to rapid thermal annealing in flowing $N_2$ at 600 – 1200 °C for 3 – 30 min. Metal composition was quantified with X-ray fluorescence (XRF), and anion composition was measured with Auger electron spectroscopy (AES). The crystal structure determined from laboratory and synchrotron X-ray diffraction (XRD) was measured as a function of the Mg–Mo–N chemical composition and at various deposition and annealing temperatures. More details about experimental synthesis and characterization are presented next.

Multiple Mg–Mo–N thin film sample libraries with different with Mg/Mo composition gradients were deposited by RF co-sputtering onto ambient-temperature (60 – 80 °C due to unintentional heating) commercial Si substrates, some with a CVD-grown $SiN_x$ layer, from 2" (50 mm) diameter Mg and Mo targets held at 30 – 60 W power. Typical depositions were performed in a mixture of flowing $Ar/N_2=6/3$ sccm in a total of 6 mTorr ($8x10^{-6}$ atm) pressure in a vacuum chamber with $10^{-7}$ Torr ($1.3x10^{-9}$ atm) base pressure for 60 – 120 min, resulting in 150 – 300 nm thin films. Thinner samples deposited for 15 – 30 min on $Si/SiN_x$ membrane substrates, and thicker samples deposited for 4 – 6 hours on quartz, were deposited for nanocalorimetry and PDF measurements, respectively. Some samples had a 50 nm thin $SiN_x$ or AlN capping layer deposited on the top of the film to prevent surface oxidation during the storage. The synthesized samples have been treated by rapid thermal annealing (RTA, ULVAC MILA-5000 series) in the 600 – 1200 °C temperature range for 3 – 30 min with a 1 min temperature ramp up and a 2 – 5 min temperature quench in flowing $N_2$.

Cation composition of the Mg–Mo–N films was measured using XRF (Bruker M4 Tornado) and quantified using XMethods software with a manual conversion from wt % to at %. The anion composition and unintentional oxygen content (if any) was determined from AES measurements using standard sensitivity factors for O and N, with Mg and Mo compositions calibrated to those of the XRF values. Nanocalorimetry measurements were performed on as-deposited Mg–Mo–N films in a $N_2$ atmosphere, with an average heating rate of roughly 10,000 °C/s. The nanocalorimeter sensor includes a Pt thin film heater deposited over a $SiN_x$, membrane which is suspended over a Si frame. The Pt heater also functions as a temperature sensor based on the temperature coefficient of resistance of Pt through calibration using the melting point of Al. A 10 nm alumina barrier was deposited between the sensor and sample using atomic layer deposition to prevent any potential reaction between sample and $SiN_x$.

Laboratory XRD patterns were collected with Bruker D8 Discover with Cu Kα radiation equipped with a two-dimensional (2D) detector and analyzed using CombIgor software,[44] with data available in NREL's HTEM DB.[45] In-situ laboratory XRD measurements were performed on a resistively heated stage in a Kapton dome under flowing $N_2$. To study the kinetic mechanism nucleation and growth during this phase transformation process, we performed analysis of the time-dependent XRD data using sigmoidal Avrami equation from the Johnson–Mehl–Avrami–Kolmogorov (JMAK) model[37] - $I(t)=1-\exp(-kt^n)$ - where I is the measured XRD intensity, $n=n_n+n_g$ are nucleation and growth components of the Avrami exponent, and $k$ is a constant. Given the characteristically sharp temperature dependence (Fig. 4a), we assume an instantaneous heterogeneous nucleation process ($n_n$ = 0). This model is commonly used to describe isothermal nucleation of growth kinetics in different areas of materials science, for example polarization domains in ferroelectric materials, most recently nitrides.[46]

High-resolution synchrotron grazing incidence wide-angle X-ray scattering (GIWAXS) measurements were performed at beamline 11-3 at the Stanford Synchrotron Radiation Lightsource (SSRL), SLAC National Accelerator Laboratory. The data were collected with a Rayonix 225 area detector at room temperature using a wavelength of λ = 0.9744 Å, a 1° or 3° incident angle, a 150 mm sample-to-detector distance, and a spot size of 50 µm × 150 µm. The diffraction images were integrated and processed with GSAS-II. Synchrotron pair distribution function (PDF) and diffraction (XRD) measurements were measured on select samples grown on quartz substrates at beamline 11-ID-B at the Advanced Photon Source (APS), Argonne National Laboratory. The data were collected in grazing incidence geometry with a wavelength of 0.1432 Å. The data were integrated using GSAS-II; PDFs were generated using xPDFsuite with $Q_{max}$ = 25 $Å^{-1}$; and PDF fitting was performed in PDFgui.[47] LeBail fits were performed on the XRD data with Topas v6 in s.g.225 (*Fm*-3*m*). The lattice parameter (*a*) and sample displacement were refined. Peak shapes were fit with a TCHZ function, with the background fit to a 10-term polynomial.

***Theoretical***

The first-principles random structure sampling method entails creating a large number of structures with: (i) random lattice vectors and atomic positions, (ii) fixed stoichiometry (e.g. $MgMoN_2$), and (iii) a fixed number of atoms (*N*=24). These are subsequently relaxed to the closest local minimum on the PES using density functional theory (DFT), and then grouped into classes of equivalent structures. The number of structures belonging to each class, or its frequency of occurrence in the structure sampling, is then taken as the size of a corresponding local minimum on the PES and used in the statistical mechanics treatment of the results. An initial pathway for the proposed mechanism for the transformation from RS to RL was computed via atom-to-atom mapping.[39] The pair distribution function (PDF) $g(r)$ for the theoretical representation of $MgMoN_2$ is directly calculated in real space.[48] More details are described next.

The first principles structure sampling, which is used for structure predictions in this work, proceeds according to the following protocol. In the first step, a large number of structures are generated with random lattice parameters (a,b,c,α,β,g) using the algorithm that ensures cation-anion coordination.[33] A total of 5,000 structure samples are constructed for $MgMoN_2$, and 2,000 structures for $MgWN_2$ and the other studied nitride materials. In the second step all random structures are relaxed, including the cell parameters and atomic positions, using density functional theory (DFT) to the closest local minimum, following the steepest descent algorithms as implemented in VASP computer code[49] within the PAW formalism using soft pseudopotential for N and normal pseudopotentials for all metals. This choice of pseudopotentials balances computational efficiency and structural accuracy. Monkhorst-Pack *k*-point grid with the subdivision determined by the VASP $R_k$=20 parameter and plane wave cutoff of 340 eV are used. Automatic *k*-point grid generation with length parameter $R_k$=20 enables consistent *k*-grid density for the random sampling, in which structures are initialized with random lattice vectors. Relaxations are considered converged when the forces on atoms are all below 0.01 eV/Å and the final pressure is below 3 kbar.

The relaxed structures are grouped into classes having the same underlying (parent) crystal structure. The classification is performed based on (a) the space group assignment of the equivalent structure in which all of the cations (Mg and Mo in case of $MgMoN_2$) are labeled the same, and (b) the first shell coordination of the atoms. For the samples grouped into classes with the same underlying structure, partition functions and corresponding configurational free energies are then evaluated as a function of temperature using energy per formula unit in the Boltzmann exponents.[35] To analyze the local cation ordering in different predicted disordered RS $MgMoN_2$ structure variants, each structure was scanned for reproduction of the *M*–*M* distances and bond angles in the cation-layered RS motif, followed by visual verification of the specific arrangement of diagonal edge-sharing *M*-centered octahedra.

To calculate pair distribution function (PDF) $g(r)$ in real space,[48] for each of the 314 RS-type structures obtained from the random structure sampling, we compute atom type pair specific PDFs $g^{S}_{\alpha\beta}(r)$ where $\alpha$ and $\beta$ denote different atom types. The ensemble averaged $g_{\alpha\beta}(r)$ is then obtained through a sum over each structure $S$ of the ensemble. The total PDF $g(r)$ is then easily obtained as a weighted sum of the partials with the weights coming from an approximation of the X-ray atomic form factors $f_{\alpha}(Q)$. The function mainly used in this work is the reduced PDF $G(r)$ as it can directly be obtained from the Fourier transform of XRD measurements:

$$G(r) = 4\pi r n_0 (g(r) - 1) \qquad (2)$$

The partials $G_{\alpha\beta}(r)$ are obtained in a similar way by just replacing $g(r)$ by the appropriate partial $g_{\alpha\beta}(r)$. More details of the PDF calculations are provided in the supplementary materials.

To compute an initial pathway for the proposed mechanism for the transformation from RS to RL, different supercell representations are chosen for the initial and final structures, and combinations of cells with maximum volume overlap are chosen for further analysis.[39] For each choice of cells, atoms are matched between the initial and final cell, solving the assignment problem using the Munkres algorithm. We start by modeling a simple transformation as if all cations have the same chemical identity, and then after the mechanism is found we re-label the atoms based on their correspondence between the two structures. In this way, we find a particular ordering in the RS phase that produces the simplest transformation based on our structure mapping. For the ordered structure, the minimum energy pathway at zero kelvin was computed using solid-state nudged elastic band (SSNEB)[50] and was considered converged when forces between images were reduced to 0.1 eV/Å, and the barriers were estimated using the climbing image method. Evaluation of upper bounds for the barrier for transformation with disorder were performed by performing site swaps between Mg and Mo in 1×1×2, 1×2×1, and 2×1×1 supercells. Energies of images along these pathways are computed without any relaxations, and therefore represent an upper bound of the energy relative to that which would be found using SSNEB. Under cation disorder conditions, the rate-determining step is the inversion of the octahedra rather than the shift of the N atoms, but the barrier associated with this step remains approximately 90 meV/atom or less. When all layers contain 50% Mo and 50% Mg (red line in Fig. S7), some of the site swaps associated with this model increase rather than decrease the energy of the RL structure to above its associated RS structure.

**Data availability**

The data used in the graphs is published alongside the manuscript. Datasets for XRD, AES, PDF, and nanocalorimetry figures can be obtained at 10.6084/m9.figshare.26344993. Crystal structure figure files can be obtained at 10.6084/m9.figshare.26345092

**Acknowledgements**

This work was authored in part at the National Renewable Energy Laboratory (NREL), operated by Alliance for Sustainable Energy, LLC, for the U.S. Department of Energy (DOE) under Contract No. DE-AC36-08GO28308. Funding was provided by the Office of Science (SC), Basic Energy Sciences (BES), Materials Chemistry program, as a part of the Early Career Award “Kinetic Synthesis of Metastable Nitrides.” (experimental results), with contribution from NSF Career Award No. DMR-1945010 (computational results). Nanocalorimeter fabrication was performed in part at the NIST Center for Nanoscale Science & Technology (CNST). Use of the Stanford Synchrotron Radiation Lightsource, SLAC National Accelerator Laboratory is supported by DOE’s SC, BES under Contract No. DE-AC02-76SF00515. This research used resources of the Advanced Photon Source, a U.S. Department of Energy (DOE) Office of Science user facility operated for the DOE Office of Science by Argonne National Laboratory under Contract No. DE-AC02-06CH11357. R.W.S. acknowledges support from the Director's Fellowship within NREL's Laboratory Directed Research and Development program. We would like to thank Craig Perkins for help with AES analysis; Nicholas Strange for assistance with SSRL data collection; Uta Ruett and Moira Miller for assistance with APS data collection; Baptiste Julien and Kei Yazawa for useful discussions. This work used high-performance computing resources located at NREL and sponsored by the Office of Energy Efficiency and Renewable Energy. Certain commercial equipment, instruments, or materials are identified in this document. Such identification does not imply recommendation or endorsement by the National Institute of Standards and Technology (NIST), nor does it imply that the products identified are necessarily the best available for the purpose. The views expressed in the article do not necessarily represent the views of the DOE or the U.S. Government.

**Author contributions**

A.Z., Y.F., C.L.R., S.R.B., O.B., D.A.L, R.W.S. conducted experimental measurements; M.J., L.W., V.S. performed theoretical calculations; A.Z. conceived the study, synthesized the materials, supervised the work, and wrote the manuscript with edits from R.W.S., M.J., L.W., V.S. and all other co-authors.

**Competing Interests**

The authors declare no competing interests.

**Supplementary information**

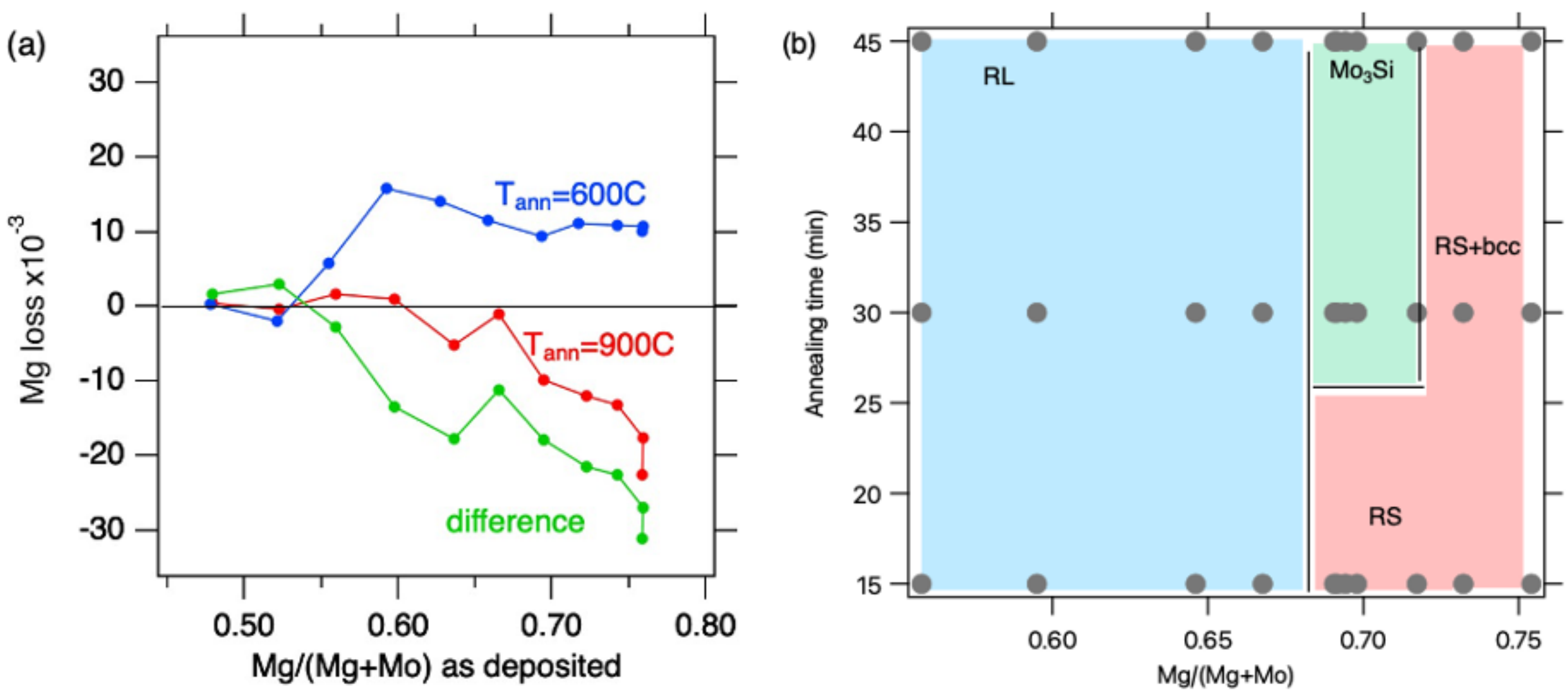

Figure S1: Chemical composition: (a) Mg loss measurements in Mg–Mo–N films as a function of Mg/(Mg+Mo) composition after 600 °C and 900 °C annealing, showing increased Mg loss up to 3% at film compositions that were initially Mg-rich; (b) Time dependence of the phase for Mg–Mo–N films annealed at 1000 °C as a function of time, showing that the material remains in the RL or RS structure until Mo starts reacting with Si substrate forming $Mo_3Si$.

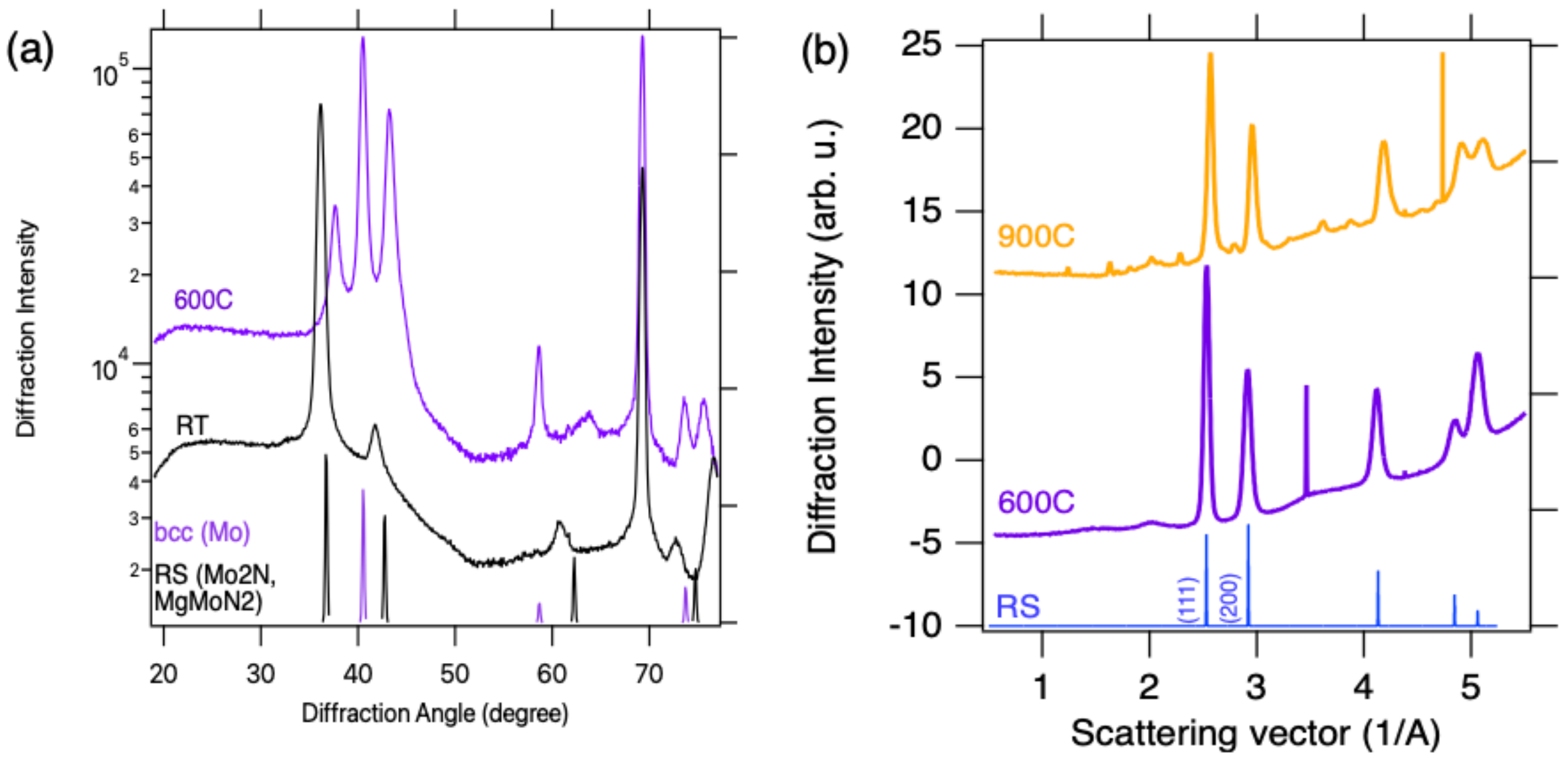

Figure S2: Long-range structure by XRD: (a) Mg–Mo–N films grow in the RS structure at both ambient (RT) and at 600 – 700 °C substrate temperatures, as measured with laboratory XRD; (b) Synchrotron GIWAXS measurement results for $Mg_3MoN_4$ annealed at 600 and 900 °C, indicating RS structure as measured at SSRL beam line 11-3.

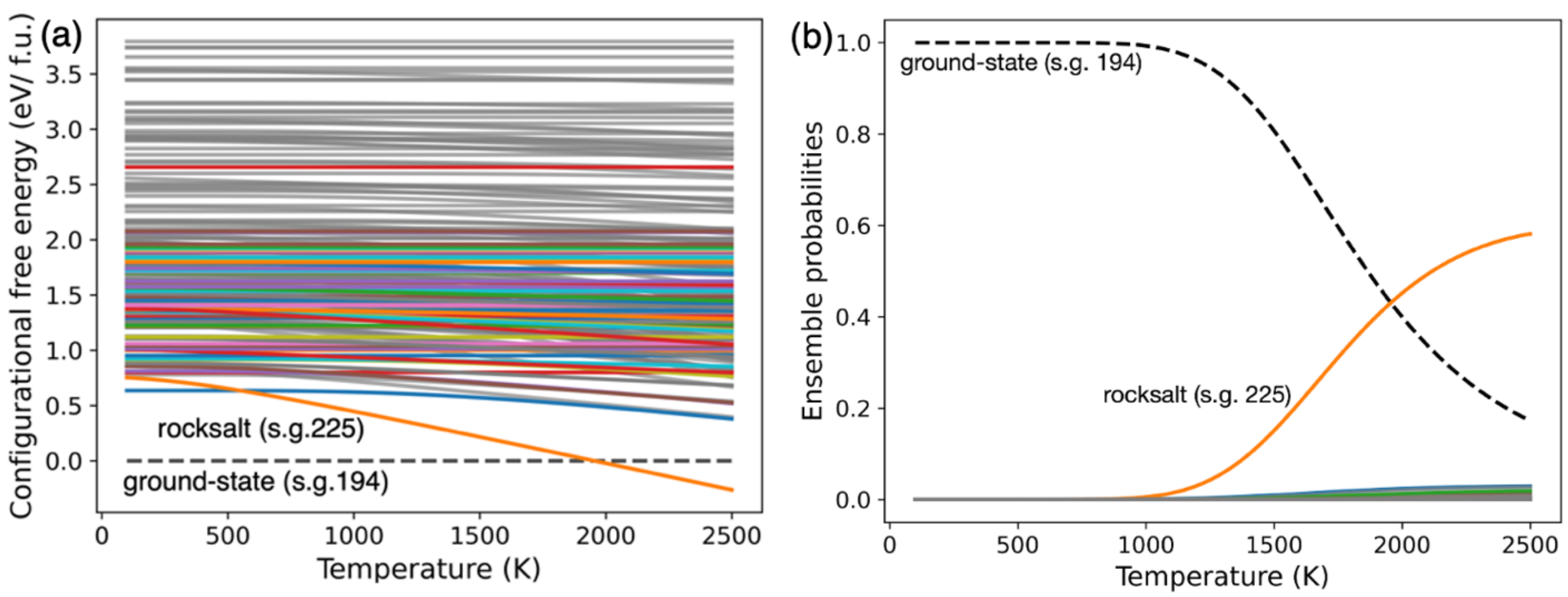

Figure S3: Polymorph sampler theoretical results: (a) Configurational free energy as a function of temperature of all structure types of $MgMoN_2$, showing that RS sg225 is the lowest energy polymorph. (b) Ensemble probabilities for the two most likely structures in thermodynamic equilibrium, showing that RS polymorph becomes dominant over the ground state RL structure above 2000 K.

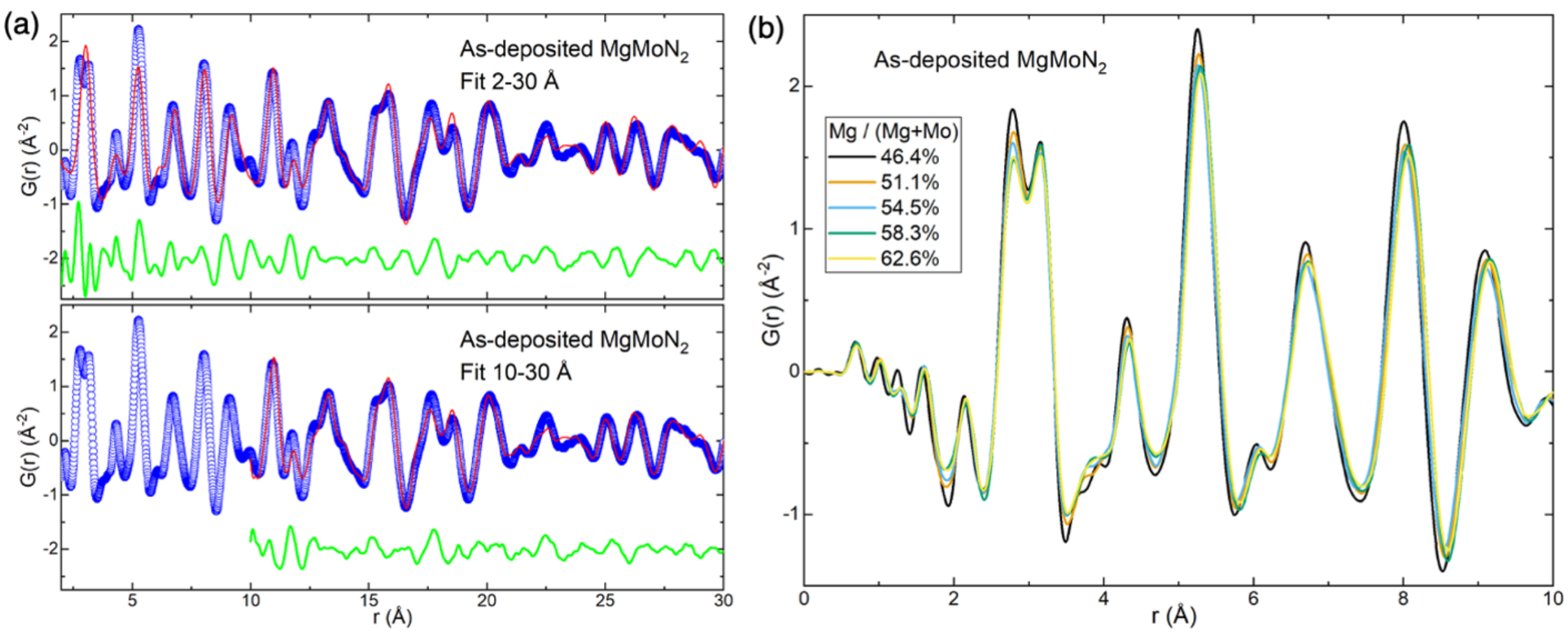

Figure S4: Short-range structure by PDF: (a) Measured PDF data (blue circles) of as-deposited $MgMoN_2$ with Mg/(Mg+Mo) = 51.1% fit with a disordered rocksalt model across the full *r*-range (2 – 30 Å, top) and a longer-length scale range (10 – 30 Å, bottom). The red line is the fit, and the green line is the difference. (b) PDF data of as-deposited $MgMoN_2$ with varying cation ratios Mg/(Mg+Mo). The data were collected at APS beamline 11-ID-B.

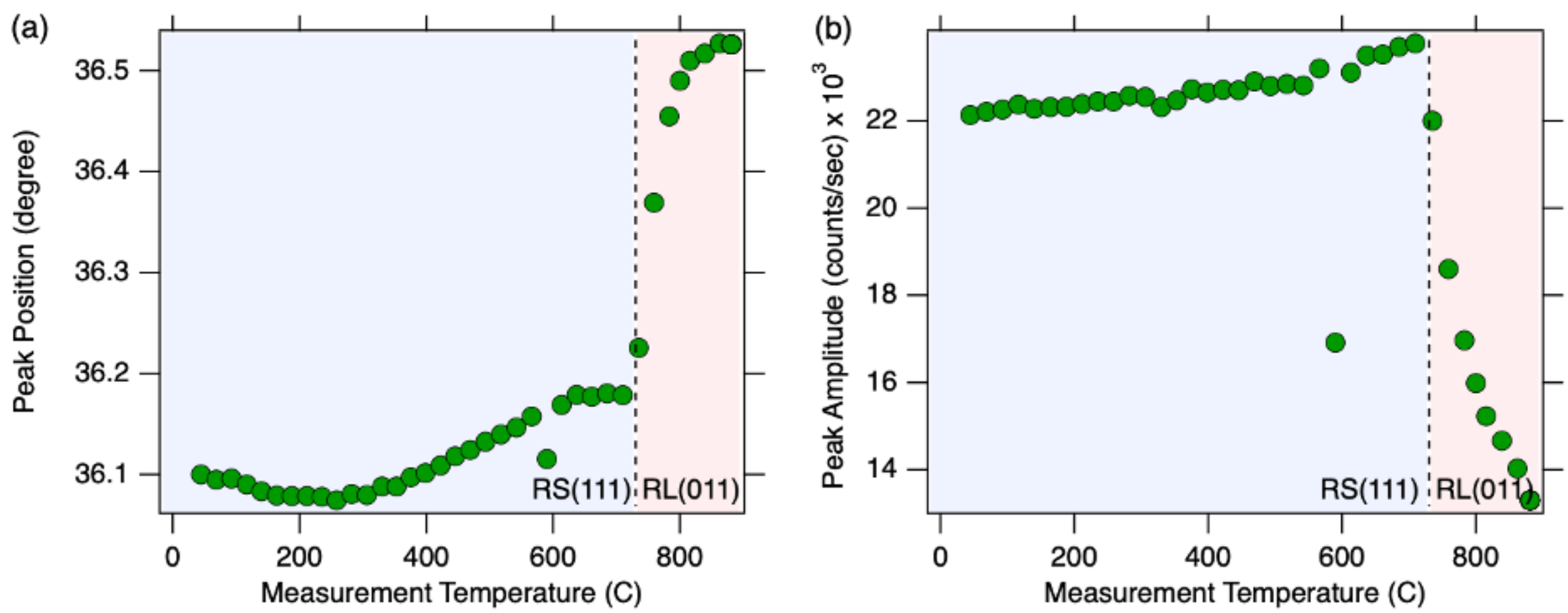

Figure S5: Temperature-dependent in-situ XRD results: (a) positions and (b) amplitudes of the RS (111) and RL (011) peaks, illustrating apparent increase in peak position and suggesting decrease in lattice volume of the RS phase above 300 °C indicative of the negative thermal expansion.

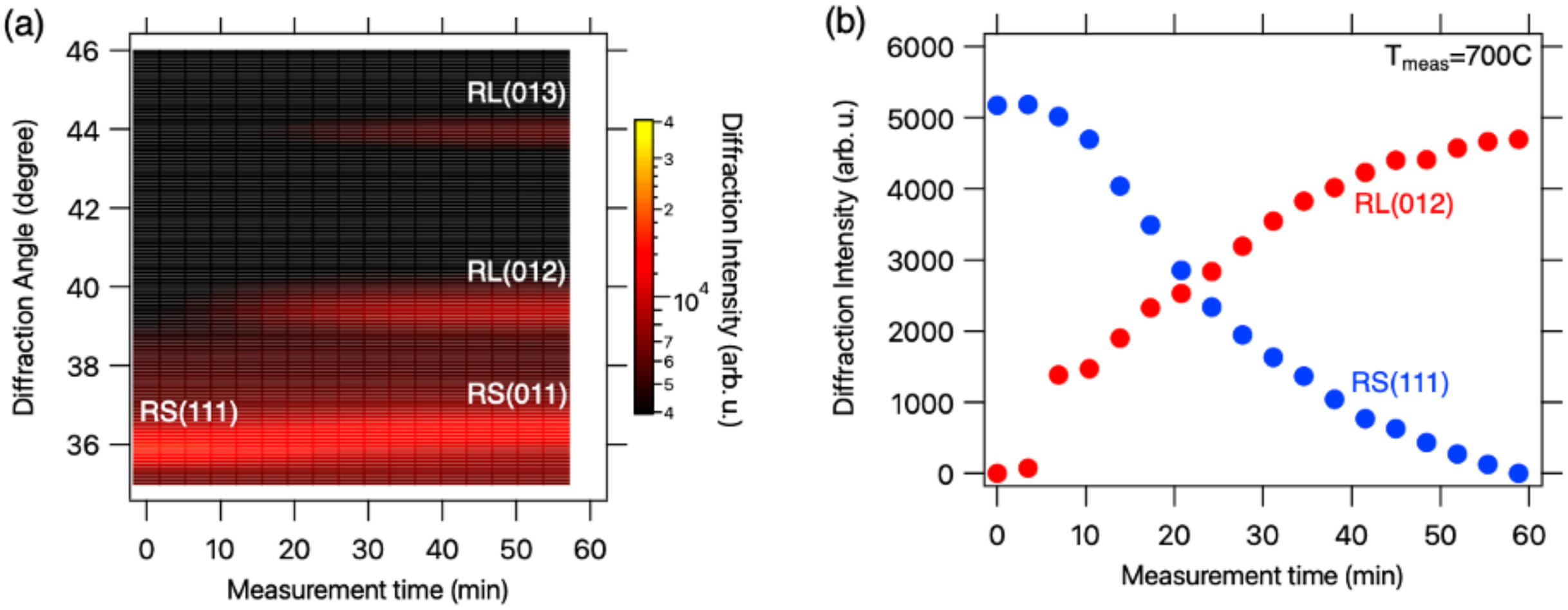

Figure S6: Time-dependent in-situ XRD results: (a) Time-dependent XRD showing evolution of the structure from RS to RL structure at 712 °C; (b) time-dependent XRD peak intensities at 712 °C measurement temperatures showing sigmoidal change described by the KAI model.

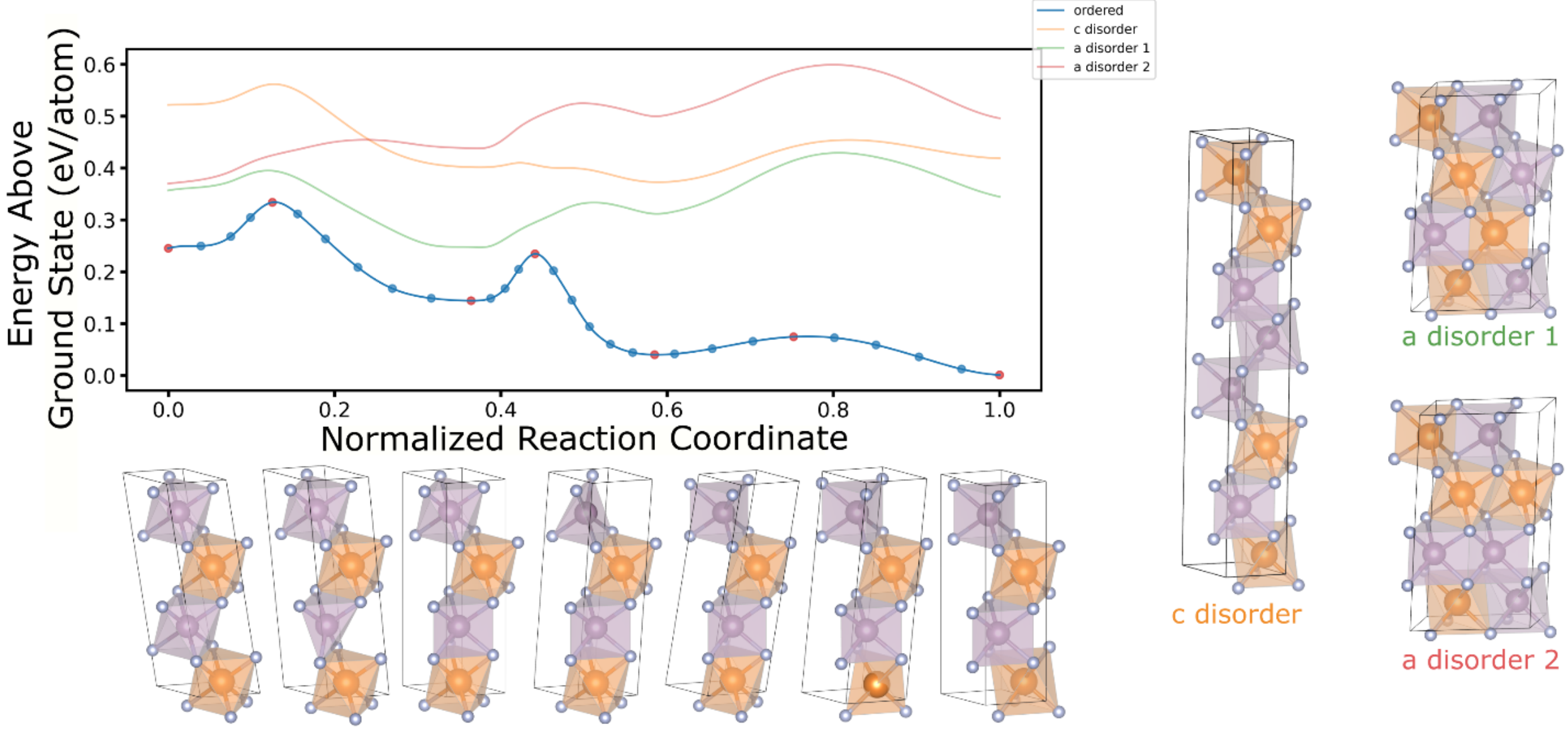

Figure S7: Nudged Elastic Band calculations: The energy profile for the transformation of the layered RS structure to the RL structure in the presence of three different versions of atomic disorder shown on the right, as compared to an ordered pathways, as calculated using SS-NEB. The structures of local maxima and minima along the ordered pathway are shown at the bottom. The rate-limiting steps of these test cases still have energy penalties on the order of 0.09 eV/atom in the cases with cation site swaps.

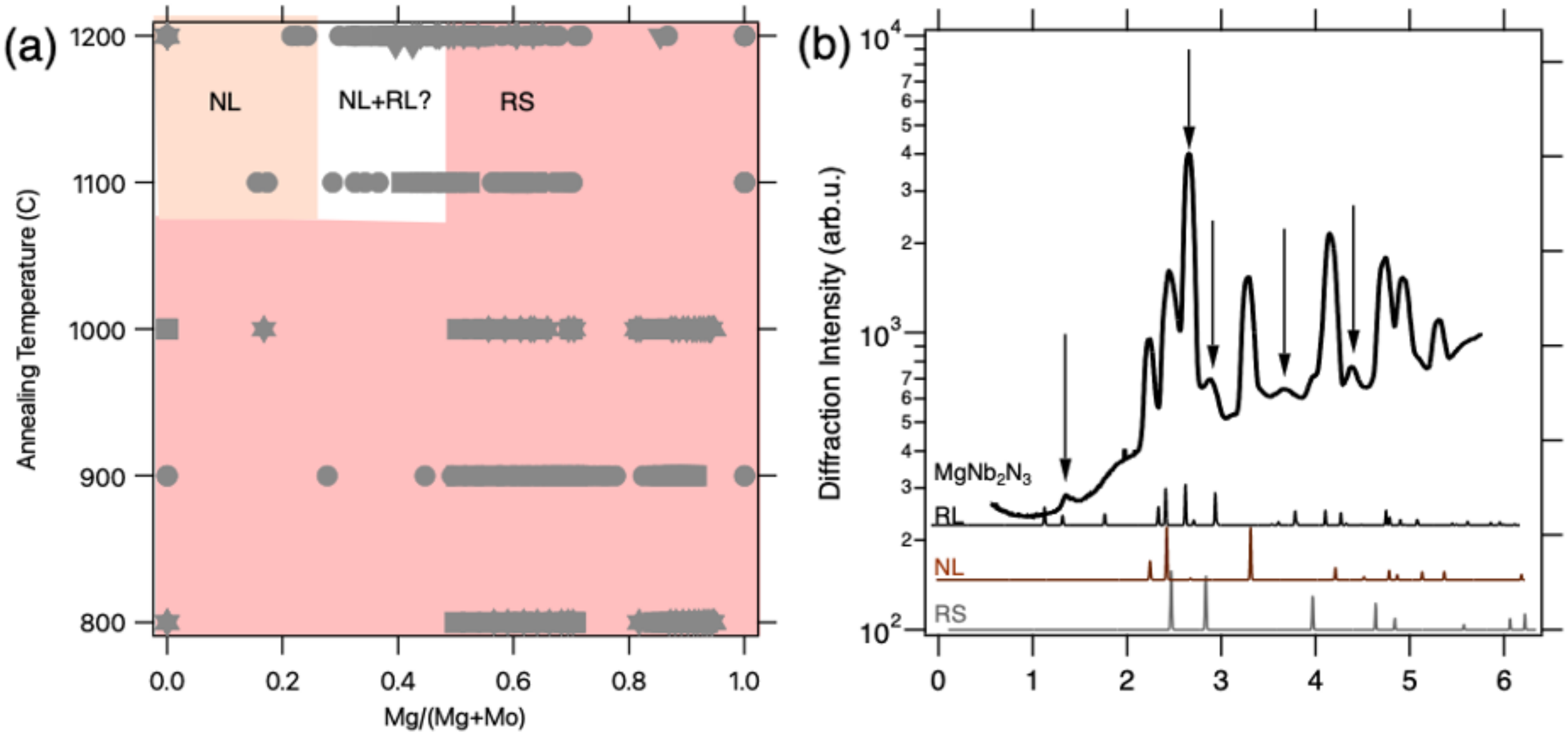

Figure S8: Synthesis results for the $MgNb_2N_3$ compound: (a) phase map of annealing results for Mg-Nb-N films, showing strong competition between RS and nickeline (NL) phases, with some minority RL phase in between. (b) A representative XRD pattern from the mixed phase region, showing a minority RL structure peak indicated by arrows, in the dominant nickeline (NL) structure of NbN phase.

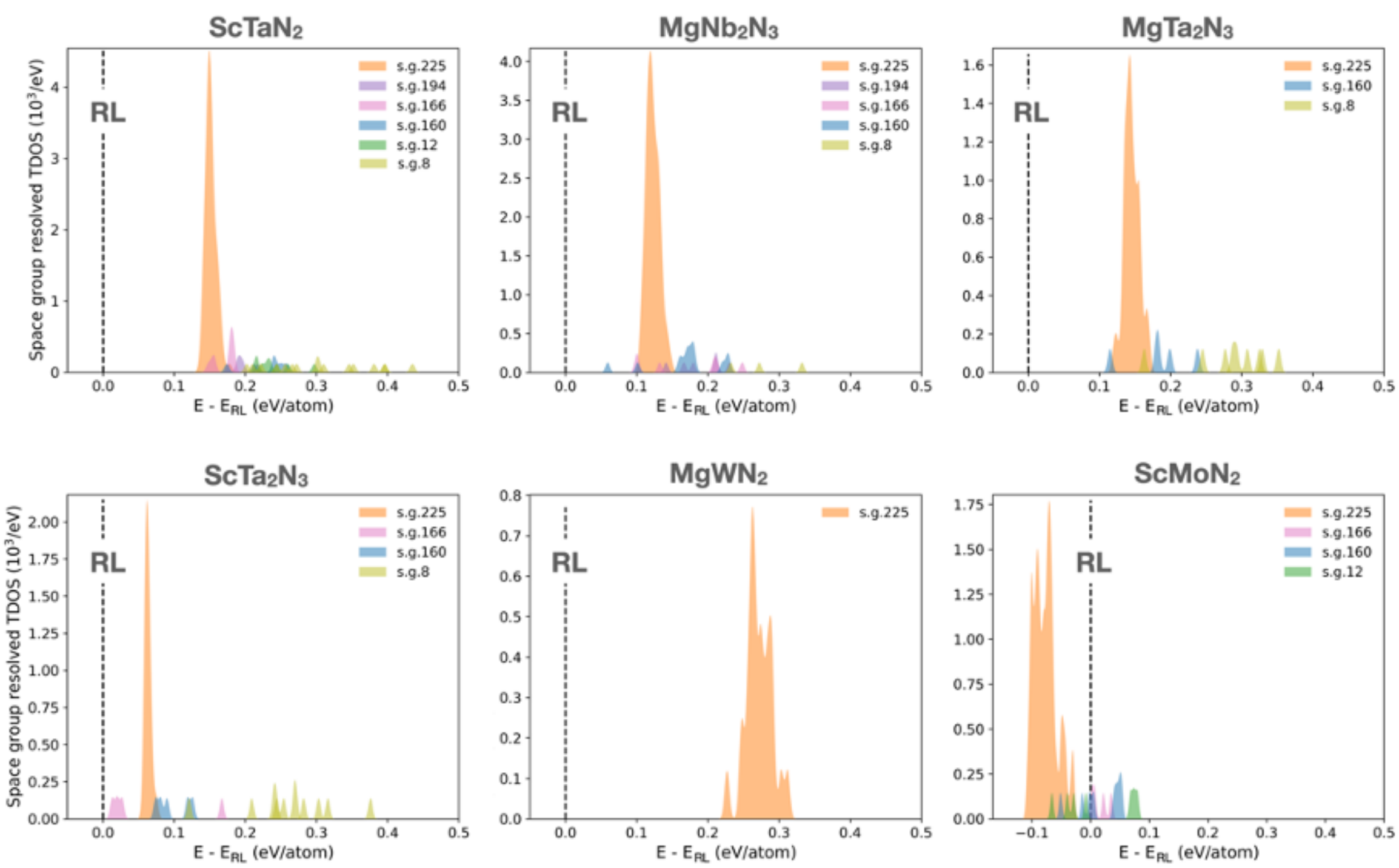

Figure S9: TDOS calculations for six candidate materials: Comparison between calculated TDOS of the disordered RS structure (s.g.225) and other predicted structures for all six calculated nitride compounds indicated in the figure, with the energies expressed relative to the RL layered structure type.

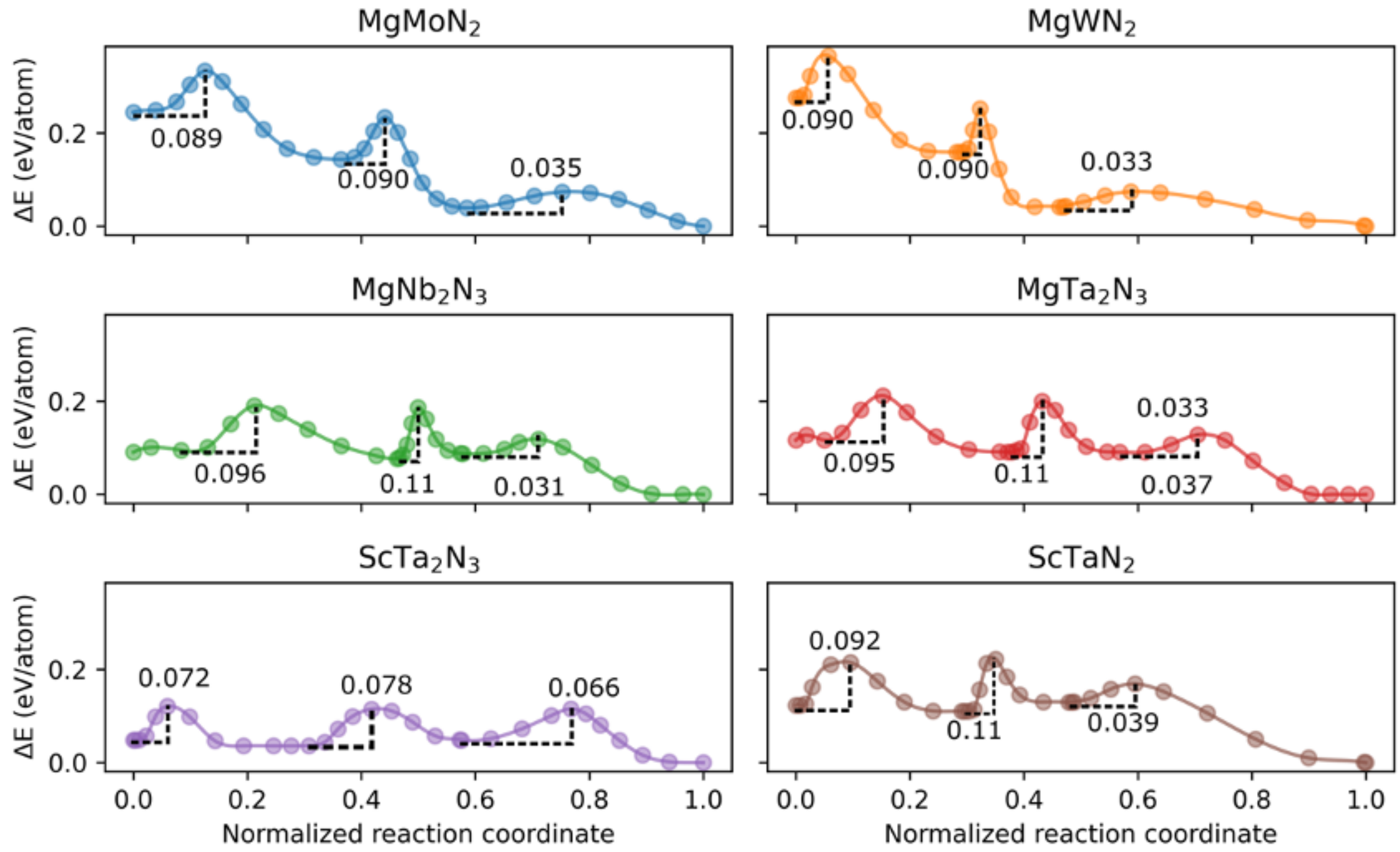

Figure S10: Barrier calculations for six candidate materials. The calculated barriers for the structural transformation from the ordered RS phase into the ordered RL phase for six calculated ternary nitride compounds using solid state NEB method.

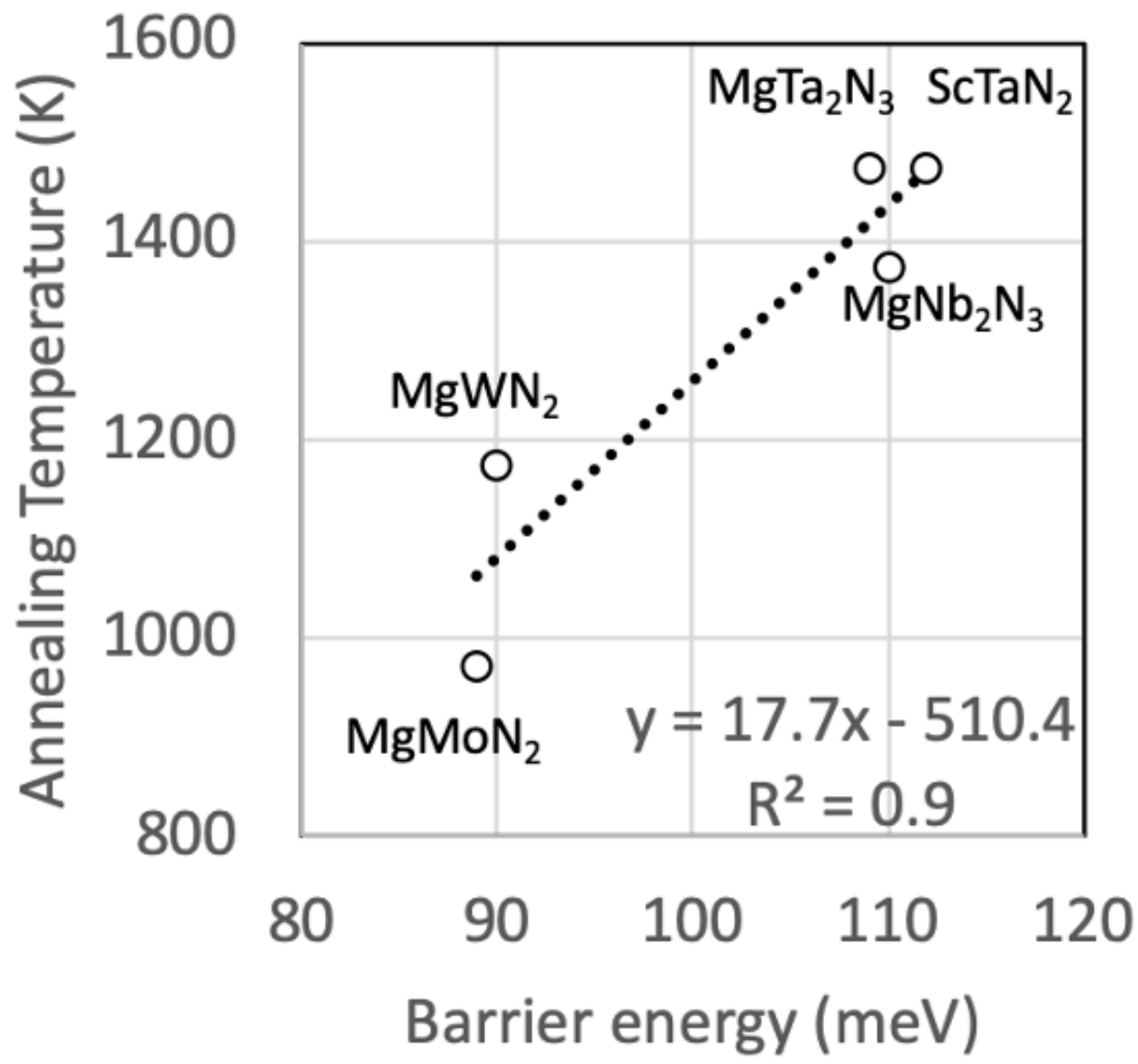

Figure S11: $T_{ann}$ vs $E_b$ correlation for 5 layered materials: Calculated transformation energy barriers $E_b$ vs measured minimal annealing temperatures $T_{ann}$ needed for the transformation from RS to RL phase.

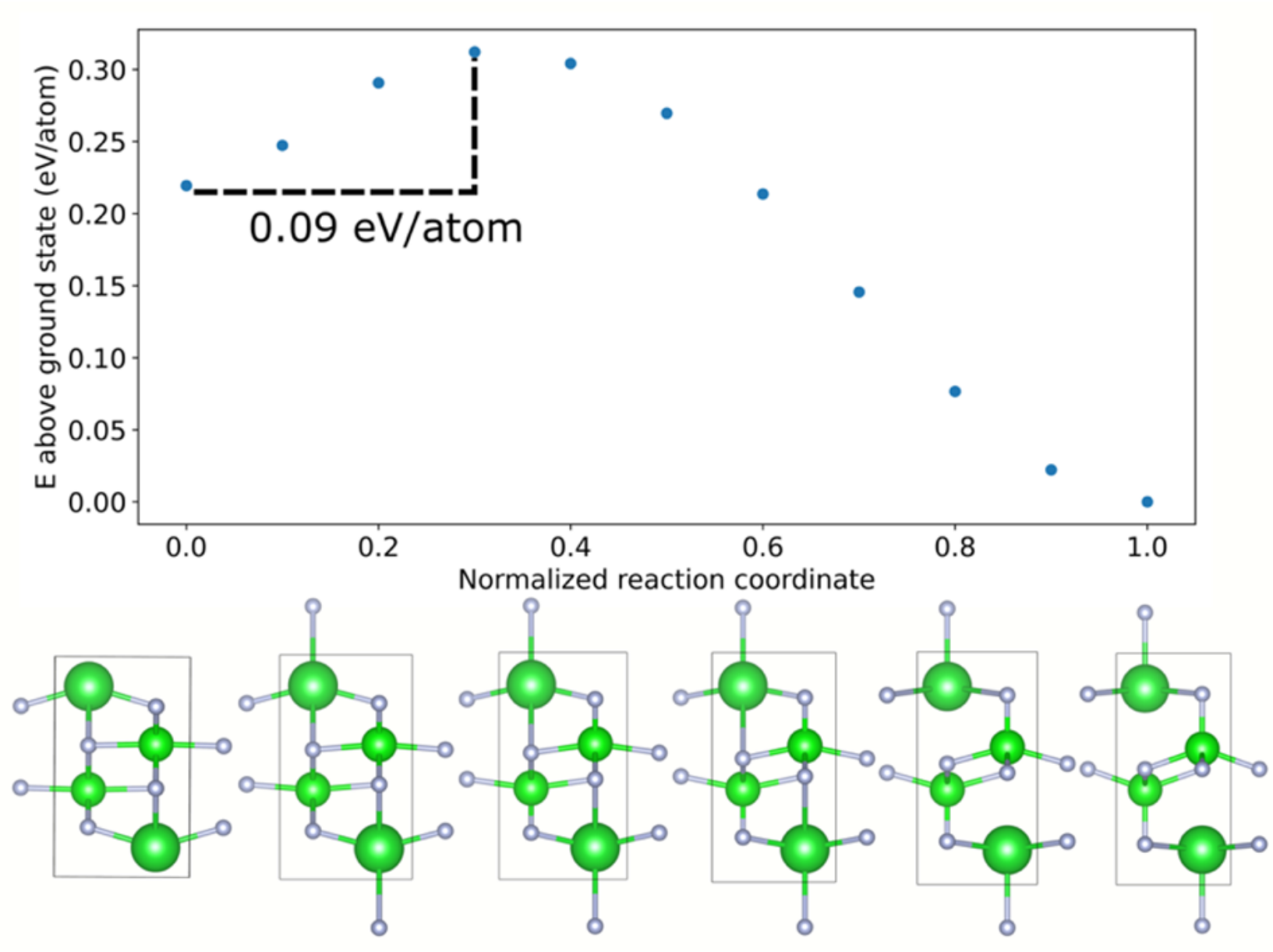

Figure S12: Calculated transformation barrier for BaZrN2: the upper bound of the energy transformation barrier for unrelaxed transformation pathway for $BaZrN_2$, from cation-ordered rocksalt-derived structure (space group *Pmmn*) to site-ordered $KCoO_2$-type structure.

Table S1: Theoretical structure sampling of ternary nitrides: Number of identified RS structures with 1-1-2 stoichiometry, and the fraction of the atoms participating in at least one cation-layered RS motifs. Similar results are also given for motifs along different axes in $MgMoN_2$.

| Composition (direction) | Total Number of Structures from DFT sampling | Number of RS-like structures | RS-like structures with identified layered RS motif | $\langle \frac{\text{cations in motif}}{cations\ in\ structure} \rangle$ |
|---|---|---|---|---|
| $MgMoN_2$ | 4827 | 314 | 314 | 99.7% |
| $MgMoN_2$ (100/110) | 4827 | 314 | 314 | 99.9% |
| $MgMoN_2$ (111) | 4827 | 314 | 312 | 88.2% |
| $MgWN_2$ | 1954 | 26 | 26 | 99.3% |
| $ScMoN_2$ | 1898 | 71 | 71 | 99.9% |
| $ScTaN_2$ | 1939 | 80 | 80 | 98.7 |

Supplementary Methods: PDF calculation

For each of the 314 RS-type structures obtained from the structure sampling, we compute atom type pair specific PDFs $g^S_{\alpha\beta}(r)$ where α and β denote different atom types:[48]

$$g^S_{\alpha\beta}(r) = \langle \frac{1}{4\pi r^2 n_0 c_\beta} \frac{dN_{\alpha\beta}(r)}{dr} \rangle_\alpha$$

Here $n_0$ is the global concentration of atoms (number of atoms/volume), $c_\beta$ is the atomic fraction of β atoms. $dN_{\alpha\beta}(r)$ represents the number of β atoms in a spherical region of thickness $dr$ at the distance $r$ from a particular α atom, the angle brackets $\langle \ldots \rangle_\alpha$ denotes an average over all $\alpha$atoms of the structure.

The ensemble averaged $g_{\alpha\beta}(r)$ is then obtained through a sum over each structure $S$ of the ensemble:

$$g_{\alpha\beta}(r) = \sum_S g^S_{\alpha\beta}(r) P^S \frac{n^S}{\langle n \rangle}$$

Where $P^S$ is the ensemble probability of structure $S$ which is taken to be $\frac{1}{number\ of\ structures}$we are working in the high temperature limit. $n^S$ is the atom concentration of structure $S$ while $\langle n \rangle$ is the average atom concentration over the ensemble.

The total PDF $g(r)$ is then easily obtained as a weighted sum of the partials with the weights coming from an approximation of the X-ray atomic form factors $f_\alpha(Q)$:

$$g(r) = \sum_{\alpha,\beta} \frac{c_\alpha c_\beta Z_\alpha Z_\beta}{\langle Z \rangle} g_{\alpha\beta}(r)$$

Where $Z_\alpha$ is the atomic number of the α atom and $\langle Z \rangle$ is the average atomic number of the system.

We note the greater sharpness of the theoretical PDF comes from the lack of vibrational and/or instrument broadening. The only broadening of the theoretical peaks arises from the diversity of the cation decorations over the RS structure produced by the structure sampling. A slight disagreement in intensity between the experimental and theoretical PDFs for certain peaks is attributed to the metal-nitrogen (*M*–N) coordination shells. This difference is likely due to a combination of slight N deficiency and the complicated grazing incidence geometry in which the experimental data were collected.